\documentclass[11pt]{article}

\let\xor\oplus
\makeatletter
\def\txtline#1{\noalign{\hbox{\strut\hskip\@totalleftmargin {#1}}}} 
\makeatother
\newcommand{\ket}[1]{|{#1} \rangle}
\newcommand{\bra}[1]{{\langle {#1}|}}
\newcommand{\braket}[2]{\langle {#1} | {#2} \rangle}
\def\<{\begin{equation}}
\def\>{\end{equation}}
\newtheorem{thm}{Theorem}[section]
\newtheorem{lemm}{Lemma}[section]

\newtheorem{defn}{Definition}[section]

\begin{document}

\title{A
Proof of the Security of Quantum Key Distribution}
\author{Eli Biham\thanks{Computer Science Department, 
Technion, Haifa 32000, 
Israel}, Michel Boyer\thanks{ DIRO, 
Universit\'e de Montr\'eal, Montr\'eal, Canada }, 
P. Oscar Boykin\thanks{ Dept. of Electrical Engineering, 
UCLA, Los Angeles, CA 90095-1594, USA}, 
Tal Mor$^\ddagger$, and Vwani 
Roychowdhury$^\ddagger$}

\date{December 2, 1999}

\maketitle

\begin{abstract}
We prove the security of quantum key distribution 
against the most general attacks which
can be performed on the channel, by an eavesdropper who has
unlimited computation abilities, and the full power allowed by the
rules of classical and quantum physics.
A key created that way 
can then be used to transmit secure messages in a way that
their security is also unaffected in the future.
\end{abstract}

\section{Introduction}

Quantum key distribution~\cite{BB84} uses the power of quantum
mechanics to suggest the distribution of a key that is secure
against an adversary with unlimited computation power. Such a task is
beyond the ability of classical information processing.
The extra power gained by the use of quantum bits (quantum two-level
systems) is due to the fact that the state of such a system cannot be
cloned.
On the other hand, the security of conventional key distribution
is based on the (unproven) existence of various
one-way functions, and mainly on the difficulty of factoring large
numbers, a problem which is assumed to be difficult for a
classical computer, and is proven to be easy for a hypothetical
quantum computer~\cite{Shor}.

Various proofs of security were previously obtained against
collective attacks~\cite{BM97a,BM97b,BBBGM}, 
and we continue this line of research here to prove the
ultimate security of quantum key distribution (QKD), against any
attack (under the conventional assumptions of QKD, as explained
below). Note that the eavesdropper is assumed to have unlimited
technology (e.g.,  a quantum memory, a quantum computer), while the legitimate users use practical tools
(or more precisely, 
simplifications of practical tools).

To prove security against such a super-strong eavesdropper
we develop some important technical tools and we reached some surprising
results: we show new {\em information versus disturbance} results,
where the power of quantum information theory is manifested in an
intuitive and clear way.
We show explicitly how the randomness of the bases,
and the randomness of the choice of test-bits provides the desired
security of QKD.
We adopt and generalize sophisticated tools invented in~\cite{BBBGM}:
a purification which simplifies Eve's states;
a bound on accessible information (using Trace-Norm-Difference
of density matrices)
which avoids any complicated optimization of Eve's possible
measurements; a connection between Eve's 
accessible information and the error-rate she induces.
We add some more simplifications (which were not required in the
analysis of collective attacks in~\cite{BBBGM}):
a reduction to a scheme
in which all qubits are used by Alice and Bob,
and a usage of the symmetries of the problem under investigation. 

Other security results/claims~\cite{Mayers96,Mayers98,LC98}
were recently given (see a few more details in 
Appendix~\ref{app:sec-proofs}).
The security result of Lo and Chau~\cite{LC98} 
uses novel techniques and is very
important, but it is somewhat limited.
The QKD protocol which is analyzed in~\cite{LC98} requires that 
the legitimate users have quantum memories and quantum computers, 
technologies which are far from being available to the legitimate
users.
The security result of Mayers~\cite{Mayers96,Mayers98} is equivalent
to ours in the sense that it proves the security of a realistic
protocol, against an unrestricted eavesdropper, and provides explicit
bounds on the eavesdropper's information.
There is no doubt that Mayers was the first to understand many of the
difficulties and subtle points related to the security issues.
The main problem with the proof of Mayers is its complexity,
and strict formality.
As a result, there was no consensus 
regarding its correctness and completeness.
Recently, Mayers' proof was confirmed by a few researchers (e.g.\cite{Shor-99})
and we believe that a consensus will soon be reached.

We follow the standard assumptions of QKD:
1) Alice and Bob share an unjammable classical channel. This
assumption is usually replaced by the demand  
that Alice and Bob share
a short secret key to be used for authenticating a standard
classical channel (hence the protocol is then a quantum key expansion
protocol).
2) Eve cannot attack Alice's and Bob's labs.
She can only attack the quantum channel and listen to all
transmissions on the classical channel.
3) Alice sends quantum bits (two level systems).

We prove the security of the Bennett-Brassard-84 (BB84)
protocol~\cite{BB84}, against any attack
allowed by the rules of quantum physics.
We prove the asymptotic security even for instances in which the error
rate in the transmission from Alice to Bob is up to 7.56\%;
this allowed error-rate can be increased much 
further by choosing fixed
codes and obtaining practical (rather than asymptotic)
security result.

\subsection{The BB84 Protocol and the used-bits-BB84 protocol}

Alice and Bob use four possible quantum states in two bases
(using ``spin'' notations, and connecting them to ``computation
basis'' notations):
(i)~~     $\ket{0_z} \equiv \ket{0}$;
(ii)~~     $\ket{1_z} \equiv \ket{1}$;
(iii)~~     $\ket{0_x}=\frac{1}{\sqrt{2}}(\ket{0}+\ket{1})$; and
(iv)~~     $\ket{1_x}=\frac{1}{\sqrt{2}}(\ket{0}-\ket{1})$.
We shall refer to these states as the BB84 states.

We prove in this paper 
the security of a simplified protocol in which only
the relevant bits are discussed (we call it the ``used-bits-BB84'').
The proof of the original BB84 protocol follows immediately, due to a
simple reduction, as we 
show in Appendix~\ref{app:used-bits}. 

Let us describe the used-bits protocol in detail, splitting it into 
{\em creating the sifted key} and {\em creating the final key
from the sifted key}.
This simplified 
protocol assumes that Bob has a quantum memory. 
\begin{enumerate}
\item[I.]Creating the sifted key: 
\item
Alice and Bob choose a large integer $n \gg 1$.
The protocol uses $2n$ bits.

\item
Alice randomly selects two $2n$-bit strings, $b$ and $i$ which are then
used to create qubits:
The string $b$ determines the basis $0\equiv z$, and $1 \equiv x$ of the
qubits.
The string $i$ determines the value (0 or
1) of each of the $2n$ qubits (in the appropriate bases).

Alice generates $2n$ qubits according to her selection, and sends
them to Bob via a quantum communication channel.

\item
Bob tells Alice when he receives the qubits.

\item
Alice publishes the bases she used, $b$; this step should be performed
only after Bob received all the qubits.

Bob measures the qubits in Alice's bases to obtain a $2n$-bit 
string $j$.

We shall refer to the resulting $2n$-bit string as the sifted key,
and it would have been the same for Alice and Bob, $i=j$,
if natural errors and
eavesdropping did not exist.

\end{enumerate}

\begin{enumerate} 

\item[II.]
Creating the final key from the sifted key:

\item

Alice chooses at random a $2n$-bit  
string $s$ which has exactly $n$ ones.
There are ${2n \choose n}$ such strings to choose from.

\item 

{}From the $2n$ bits,
Alice selects a subset of $n$ bits, determined by the zeros
in $s$, to be the test bits.
Alice publishes
the values of these test bits (given by a string $i_T$).
The values of Bob's bits on the test bits are given by $j_T$.

The other $n$ bits are the
information bits (given by a string $i_I$). 
They are used for deriving a final key via error
correction codes (ECC) and privacy amplification (PA) techniques.

[Alice shall send the ECC and PA information
to Bob, hence Bob needs correct his errors and use PA to obtain a 
key equal to Alice's].

\item
Bob verifies that the error rate $p_{test} = |i_T \xor j_T|/n $
in the test bits is lower than
some agreed error-rate $p_{allowed}$,
and aborts the protocol if the error rate is larger.

\item\label{it:bobsbits}
Bob also publishes the values of his
test bits ($j_T$). This is not crucial for the
protocol, but it is done to simplify the proof.

\item
Alice selects a linear
ECC with $2^k$ code words of $n$ bits,
and a minimal Hamming distance $d$ between any two words:
an $(n,k,d)$ code, and publishes it
along with the ECC parities on the information bits;
The strategy is that Alice announces the parity check
matrix of an ECC, i.e., $r=n-k$ parity check strings of n bits: $v_s$, $s=1,\ldots,r$.
She then announces $r$ bits which are the parities of her
string $i_I$ with respect to the parity check matrix, which
is $v_s\cdot i_I$ for all $s$. 
Bob doesn't announce anything.
The condition on the ECC is that it corrects
$t \ge (p_{allowed} + \epsilon_{rel})  n $ errors,
for some positive $\epsilon_{rel}$.
An ECC corrects $t$ errors if $d \ge 2t+1$, and thus
$d  \ge 2 (p_{allowed} + \epsilon_{rel})  n + 1  $ must be chosen.

\item
Bob performs the correction on the information bits.

\item
Alice selects a privacy amplification function (PA) and
publishes it.  The PA strategy is to publish $m$ $n$-bit strings and use
the parities of the bits masked by these strings
as the secret key.
That is she announces  
{\em privacy-amplification-strings}  $v_s$, where $s = r+1,\ldots,r+m$,
of $n$ bits each.
The final secret key bits are $v_s \cdot i$.
This strategy is similar to error correction except
that the parities are
kept secret.

The PA strings must be chosen such that the minimal distance
$\hat v$, between any string in their span and
any string in the span of their union with the ECC
parity-check-strings,
is at least $\hat v \ge 2 (p_{allowed} + \epsilon_{sec}) \ n$.
Note that, by definition, the minimal 
distance of the space spanned by the ECC
and PA strings, $d^\dagger$,
is less than the above distance, hence if we demand 
$d^\dagger \ge 2 (p_{allowed} + \epsilon_{sec}) \ n$,
the above criterion  is automatically satisfied.

\item
Bob performs the PA on the corrected information bits.
The result obtained is the final key.

\end{enumerate}

\subsection{Eavesdropping}

Eve attacks the qubits in two steps. First she lets all qubits pass
through a device that tries to probe their state.
Then, after receiving all the classical data, she measures the probe.
She can gain nothing by measuring the probe earlier, since such a
measurement is a special case of applying a unitary
operation (it is the application of a measurement gate).
Thus we can split Eve's attack into her transformation and her
measurement.
\begin{itemize} 

\item[]Eve's transformation:
The qubits can be attacked by Eve while they are in the channel
between Alice and Bob. Eve can perform any attack allowed by the laws of
physics, the most general one being any unitary
transformation on Alice's qubits and Eve's probe.
We are generous to Eve, allowing her to attack all the bits together
(in practice, she usually needs to send the preceding
qubit towards Bob before
she has access to the next one).

Without loss of generality
we assume that all the noise on the qubits,
is caused by Eve, and can be used by her in any way she likes.

\item[] Eve's measurement:
Eve keeps the probe in a quantum memory.
After Eve receives {\it all} the classical information from
Alice and Bob, including the bases of all bits, 
the choice of test bits, the test bits values, the ECC, the
ECC parities, and the PA,
she tries to guess the final key using her best
strategy of measurement.

\end{itemize}
Eve's goal is to learn as much information as possible on the final
key without causing Alice and Bob to abort the protocol due to a
failure of the test.
The task of finding Eve's optimal operation in these two steps is very
difficult. Luckily, to prove security that task need not be solved,
and it is enough to find bounds on Eve's optimal information (via any
operation she could have done).

\subsection{Security and Reliability}

The issue of the security criterion is non-trivial since
the obvious security criterion
(that Eve's information given that the test passed, is small) does
not work.

To be more precise, let ${\cal A}$ be a random variable presenting
Alice's final key, 
${\cal B}$ be a random variable presenting
Bob's final key, 
and ${\cal E}$ a random variable representing a
string in Eve's hands as result of her measurements.
Let ${\cal T}$ be a random variable presenting if the test passed or
failed.
What one would like to obtain as a security criterion is
$I({\cal A};{\cal E} \ |\ {\cal T}={\rm pass})\le A_{\rm info}
\ e^{-\beta_{\rm info} n}$
with $A$ and $\beta$ (with any subscript) positive constants.

Unfortunately~\cite{Mayers98},
the above bound is \emph{not} satisfied in quantum cryptography.
Given that the test is passed, Eve can still have
full information.  Consider the {\em swap
attack}: Eve takes 
Alice's qubits and puts them 
into a quantum memory.  She sends random
BB84 states to Bob. Eve measures the qubits she kept after learning
their bases, hence gets full information on Alice's final key.
In this case, 
Bob will almost always abort the protocol because
it is very unlikely that his bits will pass the test.  However,
even in the rare event when the test is passed,
Eve still has full information on Alice's key.
So, given
the test is passed (a rare event), information is still $m$ bits,
and the above criterion cannot be satisfied.  

In order to prove security we show that the event where 
the test is passed \emph{and} Eve obtains meaningful 
information on the key is extremely unlikely.  
This means, that if Eve tries an
attack that gives her non-negligible 
information on a final key 
she has to be extremely lucky in order to pass the test.
Formally, the security criterion is:
\begin{equation}
Prob({\rm Test\ Passes\ and\ } I_{Eve} \ge A_{\rm info}
\ e^{-\beta_{\rm info} n} ) \le
 A_{luck} \  e^{-\beta_{luck}n}
\ . 
\end{equation}
Where 
$I_{Eve}\equiv I({\cal A};{\cal E}|i_T,c_T,b,s)$
is the information Eve has on the key, after the particular
protocol values ($i_T$, $j_T$, $b$, $s$) 
are announced by Alice and Bob, and the probability
is calculated over the cases such that 
$c_T = i_T \xor j_T$ satisfies $|c_T| \le n p_{allowed}$.
Note that Alice and Bob can increase the number of bits $n$ as they
like to increase security.

We show that the final $m$-bit key is reliable:
the keys distilled by Alice and Bob are
identical except for some exponentially small probability
$A_{\rm rel} \ e^{-\beta_{\rm rel} n}$.

\subsection{Structure of the Paper}

The rest of the paper contains three main steps:
In Section~\ref{sec:attack} we reduce the problem to a simpler problem
of optimizing over all attacks symmetric to the bit values 0 and 1.
In Section~\ref{sec:infdis} we analyze the information bits
in the bases actually used by Alice and Bob,
and we prove our main {\em information versus disturbance} theorem:
the eavesdropper information on the final key
is bounded by the following probability: the
probability of error if the {\em other} bases were used
by Alice and Bob
(this probability is well
defined).
We then obtain 
in Section~\ref{sec:compl} a bound on 
$\sum_{i_T,c_T,b,s}P({\cal T}={\rm pass},i_T,c_T,b,s)\ I({\cal A};
{\cal E}|i_T,c_T,b,s)$, and prove that this bound is exponentially
small with $n$ (so that the security criterion is satisfied).
Various theorems are proven 
in the appendices.

\section{Eve's Attack}
\label{sec:attack}

In the protocol Alice sends a string $i$ encoded in the bases of her
choice $b$, and Bob measures a string $j$ using the same set of bases.
Eve prepares a probe in a known state, say $|0\rangle$.
Eve applies a unitary
transformation $U$ on 
{\it all} 
the qubits and her probe and then she
sends the
disturbed qubits to Bob, while leaving her probe in her hands.
The unitary transformation $U$ is written in the basis $b$,
$  U(\ket{0}\ket{i}) = \sum_j \ket{E'_{i,j}}\ket{j}$,
with $\ket{E'_{i,j}} $
the unnormalized states of Eve's probes if Alice
sent $\ket{i}$,
and Bob received $\ket{j}$.

Recall that the choice of $0/1$ is
random. 
As a result, 
any attack chosen by Eve can be replaced by an equivalent 
attack which is as good, with $i$ replaced by $i \oplus k$ 
and with $j$ replaced by $j \oplus k$. 
Thus, any attack chosen by Eve can also 
be replaced by an equivalent symmetric
attack which is as good (as described below).
The symmetrization does not 
change the induced error-rate as we show below.
It can improve Eve's final information on the common
secret key, and thus, 
if the optimal attack is asymmetric, there is also an
equivalent symmetric attack which is optimal.
Thus, the optimal attack can be assumed to be symmetric (WLG),
and we therefore need
to bound Eve's information only for attacks symmetric to 0/1.

The symmetrization is performed using bit-wise operations: 
Given any transformation
of Eve, Eve symmetrizes it as follows:
For each qubit $\ket{q_l}$, she creates a qubit $\ket{w_l}=H\ket{0} = 
\frac{1}{\sqrt{2}} (\ket{0_z}+\ket{1_z})$,
and performs 
a pseudo-controlled-NOT transform on the result of this
bit and $\ket{q_j}$: if $\ket{w_j}=0$: leave $\ket{q_j}$ as is.
Otherwise negate it (i.e., rotate by 180 degrees).
After the application of $U$, she performs the inverse of this
pseudo-controlled-NOT transform.
The gate  
needs to negate the bit in both
$x$ and $z$ bases.  In fact, Control-($\sigma_x\sigma_z$) 
on this ancillary qubit and the data qubit performs this transformation.
The 0/1 symmetrization ensures that the errors are independent of the
values 0 or 1 that Alice sends (in either basis).
The overall attack on all  qubits is then
described by
$U^{sym}(\ket{0}\ket{i}) = \sum_j \ket{E_{i,j}^{'sym}}\ket{j}$
with $E'^{sym}$ which can be written using $E'$ as follows:
\begin{equation} \label{symmetry-eq}
\ket{E_{i,j}^{'sym}}=\frac{1}{\sqrt{2^{2n}}}
\sum_{m}(-1)^{(i\xor j)\cdot m}
\ket{m}\ket{E'_{i\xor m,j\xor m}}
\ .
\end{equation}

To prove that the $0/1$ symmetrization does
not change the average error-rate is obvious
since Eve 
can always project onto one particular $m$ (and destroy the symmetry)
by measuring $|m\rangle$, and any such projection leads to the same
attack (up to a shift of $i$ and $j$ by $m$).
It is also obvious that the symmetric attack cannot be worse
(for Eve)
in terms of Eve's information (for the same reason).
Clearly, it can only increase Eve's information since she does not
have to measure $m$ but can also do other things.

Later on Eve obtains {\em all} classical information sent by
Alice and Bob.
Eve learns $b$ (the bases) and $s$ (which bits are the test bits and
which are the information bits).
She also learns the values of the test bits $i_T$ and $j_T$.
We also use $i_I$ and $j_I$ 
to denote the values of the information bits.
Once the additional
data regarding the bases and the values of the
test bits is given to Eve,
this data modifies her probes' states.
We define $|\psi_{i_I}\rangle $ 
to be the state of
Eve+Bob if Alice chose a bases $b$, an order $s$,
and values $i_T i_I$, Eve's attack is $U^{sym}$, and Bob
received $j_T$ in his measurement on the test bits.
Formally, 
\begin{equation}   |\psi_{i_I}\rangle =
\frac{1}{\sqrt{P(j_T|i_T,i_I,b,s)}}
  [\bra{j_T} ] U^{sym}
    [(\ket{0})_{Eve}(\ket{i_T}\ket{i_I})_{Alice}] 
\end{equation}	
We define $\ket{E_{i_I,j_I}}$, Eve's states {\em for a given}
classical data regarding ($i_T,j_T,s,b$), by writing 
$|\psi_{i_I}\rangle = 
	\sum_{j_I}
	   \ket{E_{i_I,j_I}}\ket{j_I}
 $.
If $E_{ij}$ is obtained from an attack $E'_{ij}$
and for that attack, $E_{ij}^{sym}$ is obtained from the 
symmetric attack $E_{ij}^{'sym}$, then $E_{ij}$ also satisfies 
Eq.~\ref{symmetry-eq} with $E$ replacing $E'$. 
See a proof in Appendix~\ref{app:symmetry}.

\section{Information Versus Disturbance}
\label{sec:infdis}

In this section we analyze the information bits alone (for a given
symmetric attack $U^{sym}$, a given input $i_T$ and outcome $j_T$ 
on the test bits, and given bases $b$ and choice of test bits
$s$). Our result here applies for any $U^{sym}$, hence in
particular for the optimal one.
The optimization over Eve's measurement is avoided by using the fact 
that trace-norm of the difference of two density matrices provides 
an upper bound on the accessible information one could obtain
when having the
two density matrices as the possible inputs.

\subsection{Eve's State}\label{subsec:Eve-states}

When Alice sends a state $\ket{i_I}$ for the information bits
(written in the basis actually used by her and Bob for these bits),
the state of Eve and Bob together,
$|\psi_{i_I}\rangle = 
	\sum_{j_I}
	   \ket{E_{i_I,j_I}}\ket{j_I}
$ 
is fully determined by Eve's
attack and by the data regarding the test bits.
Eve's state in that case  is fully
determined by tracing-out Bob's subsystem $\ket{j_I}$
from the
Eve-Bob state, and it is
\[\rho^{i_I}=  \sum_{j_I} \ket{E_{i_I,j_I}}\bra{E_{i_I,j_I}} \ ,\]
calculated given $i_T$ and $j_T$.
This state in Eve's hands is a mixed state.  

\subsection{Purification and a related Orthogonal Basis}

We can ``purify'' the state while giving more information to Eve by
assuming she keeps the state
\[ \ket{\phi_{i_I}} = \sum_{j_I}\ket{E_{i_I,j_I}}\ket{i_I\xor j_I} \]
where we introduce another subsystem for the purification.

This state is at least as informative to Eve as
$\rho^{i_I}$ is.  This is because the density matrix is exactly the
same if Eve ignores the $i\xor j$ register of $\phi$.
Thus, any information Eve can obtain from her mixed state is bounded by the
information she could get if the purified state was available to her.

Since we deal here only with the information bits, we can drop the
subscript ${}_I$ when there is no risk of confusion, and we write:
$ \rho^{i}=  \sum_{j} \ket{E_{i,j}}\bra{E_{i,j}} $, and
$  \ket{\phi_{i}} = \sum_{j}\ket{E_{i,j}}\ket{i\xor j} $.
We shall retain the index when we consider both Information and Test bits.

We define an orthogonal basis $\ket{\eta}$, and show that
it is possible to compute a bound on
Eve's information on the information bits, once the purified states are
written in this basis.
\begin{defn} \em
\begin{eqnarray*}
\ket{\eta_{i}}&=& \frac{1}{2^n}\sum_{l} (-1)^{i\cdot l}
\ket{\phi_{l}} \ ; \
d_{i}^2 = \braket{\eta_{i}}{\eta_{i}} \ ; \
\hat\eta_{i} = \eta_{i} / d_{i}
\end{eqnarray*}
\label{eta-def}
\end{defn}
Using the above definitions and $(1/2^n)\sum_l (-1)^{(i \xor j)
\cdot l} = \delta_{ij}$,
Eve's purified state can be rewritten
as:
\begin{equation}\label{eq:phi}
\ket{\phi_i}=\sum_l(-1)^{i\cdot l}\ket{\eta_l}
\end{equation}

\begin{lemm} \label{basis-eta}
\em
$\eta_i$'s form an orthogonal  basis, i.e.,
$\braket{\eta_k}{\eta_l}=0$ for $k\ne l$.
\end{lemm}
{\bf Proof:} \hspace{.5cm}
See Appendix~\ref{app:eta-basis}.
It should be noted the above lemma is the only place where $0/1$ symmetry
is explicitly made use of.

\subsection{Eve's State and Probability of Errors
Induced on Information Bits}

In this subsection we first show that the probability
of any error string Eve would
have induced if the conjugate basis was used for the
information bits, is a simple
function of $d_{i}$'s (of Definition~\ref{eta-def}).

For any attack
$P(j_I=i_I \xor c_I \ | \ i_I,i_T,j_T,b,s) =
\braket{E_{i_I,i_I\xor c_I}}{ E_{i_I,i_I\xor c_I}}$, and thus,
the error distribution in the information bits is
\[ P(c_I \ | \ i_T,j_T,b,s)
\stackrel{\Delta}{=} \frac{1}{2^n} \sum_{i_I}
  P(\hbox{$\ket{j_I}=\ket{i_I\xor c_I}\ {\rm given}
  \ i_I,i_T,j_T, b, s$}) =
\frac{1}{2^n}\sum_{i}\braket{E_{i_I,i_I\xor c_I}}{E_{i_I,i_I\xor c_I}}
                            .\]
the average probability of an error syndrome $c$ for the information
bits (when the
test bits, basis and sequence are given).

Due to the linearity of quantum mechanics,
given Eve's attack in one basis
we can write Eve's attack in any other basis, and in particular, in a
basis where the $x/z$ bases of each information qubit
are interchanged. Let an input string be the same as the original,
but with the bases of the information bits switched, i.e.,
$|i_T,i_I^o\rangle$, and let the output bits be
$|j_T,j_I^o\rangle$ in that switched bases.
If Alice and Bob used the conjugate basis for each of the information
qubits (while using the same bases as before for the test
bits), and Eve used the same attack, then the error distribution
in the conjugate basis (over the information bits) is
\[ P(c_I^o|i_T,j_T,b,s)
\stackrel{\Delta}{=} \frac{1}{2^n} \sum_{i_I^o}
  P(\hbox{$\ket{j_I^o}=\ket{i_I^o\xor c_I^o}\ {\rm given} \
  i_I^o,i_T,j_T,b,s$}) .\]

The following lemma shows that the probability of
an error syndrome $c$,
if the conjugate bases were used, equals
the coefficients $d_c$ when writing the
purification of Eve's states in the basis $|\eta_c\rangle$.
\begin{lemm} \label{d-sqr-is-p-err}
\begin{equation} \label{info-vs-dis}
P(c_I^o \ {\rm given} \  i_T,j_T,b,s) = d_{c_I}^2
\ .
\end{equation}
\end{lemm}
{\bf Proof:} \hspace{5mm} See Appendix~\ref{app:d-sqrt-is}.

This is the only place where the properties of quantum mechanics are
used, and this result is later on
translated into an {\em information versus
disturbance} result.
The rest of the proof is based on probability theory and information
theory.

\noindent
\subsection{Bounds on Eve's Information}\label{subsec:bounds}

In this subsection we improve upon a result based on~\cite{BBBGM}.
Eve's information on a particular bit of the final key
(even if all other
bits of the final key are given to her) is bounded.
We take into consideration the error-correction data that is given to
Eve, and we do it more efficiently than in~\cite{BBBGM}, hence we
obtain a much better threshold for the allowed error-rate.

Let us first discuss one-bit final key, defined to be the parity of
substring of the input $i$.  The substring is defined using a mask
$v$, meaning that the secret key is $v \cdot i$
(so $v$ tells us the subset of bits whose parity is the final
key). See Appendix~\ref{app:ECC} for more formal explanation of ECCs.
Eve does not know $i$, but she learns
the error correcting code ${\cal C}$
used by Alice and Bob as well as $v$ and the
parity bits
$\xi$ sent by Alice to help Bob correct the sequence he received.
All the possible inputs $i$ that have the correct parities
$\xi$ for the code ${\cal C}$ form a set denoted ${\cal C}_\xi$.

When the purification of Eve's state is given by $\ket{\phi_{i}}$
the density matrix is
$\rho^{i}=\ket{\phi_{i}}\bra{\phi_{i}}$.
In order to guess the key
${\hat b} = v \cdot i$, Eve must now distinguish between
two {\em ensembles} of states: the ensemble of (equally likely)
states $\rho^i$ with 
$i \in {\cal C}_\xi$ and key ${\hat b} = v \cdot i =0$,
and the
ensemble of (equally likely)
states $\rho^i$ with $i \in {\cal C}_\xi$ and
key ${\hat b}= v\cdot i=1$.
For ${\hat b}\in \{0, 1\}$ these ensembles are represented
by the following density matrices:
\begin{eqnarray*}
\rho_{\hat b}&=&\frac{1}{2^{n-(r+1)}}
\sum_{\stackrel{i \in {\cal C}_\xi}{i\cdot v={\hat b}}} \rho^i
\end{eqnarray*}
and Eve's goal is to distinguish between them.
A good measure for their distinguishability is the optimal mutual
information (known as the accessible information) that one could get
if one needs to guess the bit $b$ by performing an optimal measurement
to distinguish between the $\rho_b$.
We call this Shannon Distinguishability ($SD$) 
to emphasize that it is a
distinguishability measure, and $SD \equiv 
{\rm opt} \{ I({\cal A}_j;{\cal E}|i_T,j_T,b,s) \}$ 
where the optimization 
is over all possible measurements.  

In the same way that $v$ acts as a mask and the secret bit is
$v \cdot i$, the error-correction data also acts as masks:
the $r$ ``parity-check strings'' $v_1, v_2, \ldots v_r$,
and the parities: $\{v_1 \cdot i, v_2 \cdot i, \ldots v_r \cdot i\}$
are given to Eve. Let us assume (WLG) that these parity-check strings
are linearly independent. Eve also knows the parity of
any linear combination of the $r$
parity strings, e.g., $(v_1 \oplus v_2) \cdot i$.
As result, a total of $2^r$ parity strings and parity bits
are known to Eve.
Let us take $s$ to be an index running from $0$ to $2^r - 1$,
so we call the set of all these $2^r$ parity strings
$ S_{\mathbf s}$, and
$v_s \in S_{\mathbf s}$ means that $v_s$ is in this set.

Let $\hat v$ be the minimum Hamming distance between
$v$ and any (error correction) parity string $v_s$.
[The minimal Hamming weight of $v\xor v_s$ when the minimum is over
all strings $v_s \in S_{\mathbf s}$].
Then, for Eve's purified states
$\ket{\phi_{i}} = \sum_l (-1)^{i\cdot l} d_l\ket{\hat \eta_l}$,
we obtain that

\begin{lemm} \em \label{sd-lemm}
The Shannon distinguishability between the parity 0 and
the parity 1 of the
information bits over any PA string, $v$, is bounded above
by the following inequality:
\begin{equation}
SD_v\le \alpha +
          \frac{1}{\alpha}\sum_{|l|\ge\frac{\hat v}{2}}d_l^2\ 
\  ,
\end{equation}
where $\alpha$ is any positive constant, and $SD_v$ is the optimal
mutual information that Eve can obtain regarding the parity bit
defined by the string $v$ (given the test and unused bits).
\end{lemm}
{\bf Proof:} \hspace{5mm} See Appendix~\ref{app:SD-vs-d}.

This gives an upper bound for Eve's information about the bit
defined by this privacy amplification string $v$.
In order to prove security in case of $m$ bits in the final key,
we start by proving security of each bit when we assume
that Eve is given the ECC information and in addition, she is also
given the values of all the other
bits of the key.  This is like using a code with $r+m-1$
independent parity check strings, or like using less code words.
Since $r$ does not appear in the above bound, replacing $r$ by $r+m-1$
leaves the same result as before,
$SD_v\le  \alpha +
          \frac{1}{\alpha}\sum_{|l|\ge\frac{\hat v}{2}}d_l^2\
$,
as a bound on Eve's information on (any) one bit of the final key
(but probably causes a decrease in $\hat{v}$).

\subsection{Eve's Information versus the Induced Disturbance}

We have already shown in Eq.(\ref{info-vs-dis}) that
$P(c_I^o \ {\rm given} \  i_T,j_T,b,s) = d_{c_I}^2$.
Thus,
\begin{equation}
SD_v\le \alpha +
          \frac{1}{\alpha}\sum_{|c_I|\ge\frac{\hat v}{2}}
P(c_I^o|i_T,j_T,b,s)\
\  .
\end{equation}
This equation bounds the information
of Eve using the probability of the
error syndromes in the other basis, and it
completes the ``information versus disturbance'' result of our
proof.
Previous security proofs (for simpler attacks),
such as~\cite{FGGNP,BM97b,BBBGM} are also based
on various ``information versus disturbance'' arguments, since
the non-classicality of QKD is manifested via such arguments.

The result is expressed using classical terms: Eve's information is
bounded using the
probability of error strings with large Hamming weight.
If only error strings with low weight have non-zero probability,
Eve's information goes to zero.
Such a result is a ``low weight'' property and it resembles
a similar result with this name which
was derived by Yao~\cite{Yao95}
for the security analysis of quantum oblivious transfer.
Henceforth we no longer
concern ourselves with the delicate
issues of quantum mechanics.

{}From this point on we want to use standard information
theory and probability
notations.
Shannon Distinguishability is the optimal
mutual information between Eve's
bits (${\cal E}$) and Alice's $j^{th}$ bit (${\cal A}_j$)
(when all other PA bits are given together with the ECC data
and test data).
Therefore,
$I({\cal A}_j;{\cal E}|i_T,j_T,b,s) \le
 \alpha +
          \frac{1}{\alpha}\sum_{|c_I|\ge\frac{\hat v}{2}}
P(c_{I}^o|i_T,j_T,b,s)\ $.

When summing 
over the $m$ bits of the key, the total information
Eve receives on the final $m$-bit key is bounded by
\begin{equation}\label{eq:inf}
I({\cal A};{\cal E}|i_T,j_T,b,s) \le
m \left(
\alpha+\frac{1}{\alpha}\sum_{|c_I|\ge\frac{\hat v}{2}}
P(c_{I}^o|i_T,j_T,b,s)\ \right)
\ ,
\end{equation}
as explained in Appendix \ref{app:m-bits}.

If  
$\alpha = \sqrt{\sum_{|c_I|\ge\frac{\hat v}{2}}
P(c_{I}^o|i_T,j_T,b,s)\ }$,
then  
$ I({\cal A};{\cal E}|i_T,j_T,b,s) \le
2m \sqrt{\sum_{|c_I|\ge\frac{\hat v}{2}}
P(c_{I}^o|i_T,j_T,b,s)\ }$,
however, to derive the security criterion 
we need not fix $\alpha$ yet.

\section{Completing the Security Proof}
\label{sec:compl}

In this section we analyze the attack on the test and information
together
$  U^{sym}
    [(\ket{0})_{Eve}(\ket{i_T}\ket{i_I})_{Alice}] \equiv
	\sum_{j_T,j_I}
	   \ket{E'_{i_T,i_I,j_T,j_I}}\ket{j_T}\ket{j_I}
$.
For these states, we will bound a weighted average of Eve's information:
$\sum_{i_T,c_T,b,s}P({\cal T}=pass,i_T,c_T,b,s)\ 
I({\cal A};{\cal E}|i_T,c_T,b,s)$.
We show that the above bound is small and hence that
security is achieved.
Note that $c_T$ replaces $j_T$ from this point forward (when $i_T$ 
is given).  Recall that
$c_T=i_T\xor j_T$, so once $c_T$ is known $j_T$ is uniquely given.
We define $C_I$ and $C_T$ to be the random variables getting
the values $c_I$ and $c_T$ respectively.

\subsection{Exponentially-Small Bound on Eve's information}

We generalize here previous proofs~\cite{BMS96,BM97a,BBBGM} that
information on parity bits is exponentially small, 
to be applicable for the joint attack.


The maximum error rate that still passes the test is $p_{allowed}$
(or $p_a$).
Also recall that ${\cal T}$ denotes the random variable for the test.
Making use of Eq.~\ref{eq:inf} we get:
\begin{lemm} \em \label{sec1-lemm}
\begin{eqnarray*}
\sum_{i_T,c_T}P({\cal T}={\rm pass},i_T,c_T| b,s)
I({\cal A};{\cal E}\ |\ i_T,c_T,b,s)
\le m \left(\alpha+\frac{1}{\alpha}\ 
P[(\frac{|C_I^o|}{n}>\frac{\hat{v}}{2}) 
\cap
(\frac{|C_T|}{n}\le p_{allowed})|b,s]\right)
\end{eqnarray*}
\end{lemm}
{\bf Proof: } \hspace{2mm} The Proof is in Appendix~\ref{exp-bound1}.

For an $\epsilon$ (called earlier $\epsilon_{sec}$)
 such that $\hat{v}\ge 
2n(p_{allowed}+\epsilon)$ we get
the following bound:
\begin{eqnarray*}
\sum_{i_T,c_T}P({\cal T}={\rm pass},i_T,c_T| b,s)
I({\cal A};{\cal E}\ |\ i_T,c_T,b,s)
&\le&m\left(
\alpha+\frac{1}{\alpha}P[(\frac{|C_I^o|}{n}>p_{a}+\epsilon) \cap
(\frac{|C_T|}{n}\le p_{a})|b,s]\right)
\end{eqnarray*}

Thus far, there is nothing that causes the bound on the right hand
side to be a small number.
The result above is true even if Eve is told in advance the bases of
Alice and Bob (the string $b$),
or if she is told in advance which are the test bits
and which are the used bits (the string $s$),
two cases in which Eve easily obtains full information.

Only Eve's lack of knowledge regarding the random $b$ and $s$ provides
an exponentially small number at the right hand side.
Since Eve must fix her attack \emph{before} she knows the basis or
order, we compute the average information for a fixed attack
over all bases and orders.
This averaging has the fortunate side effect of
removing the conjugation on $C_I$:
\begin{lemm} \em \label{sec1-lemm2}
\begin{equation}
\sum_{i_T,c_T,b}P({\cal T}={\rm pass},i_T,c_T,b\ |\ s)
I({\cal A};{\cal E}\ |\ i_T,c_T,b,s)
\le m\left(\alpha + \frac{1}{\alpha}
	 \frac{1}{2^{2n}}\sum_b P[(\frac{|C_I|}{n}>p_{a}+\epsilon) \cap
	(\frac{|C_T|}{n}\le p_{a})|b,s]\right)
\end{equation}
\end{lemm}

{\bf Proof: } \hspace{2mm} The Proof is in Appendix~\ref{exp-bound2}.

By averaging over all values of 
the order $s$, and assigning a value to the free
parameter $\alpha$ we get:
\begin{lemm} \em \label{sec1-lemm3}
\begin{equation}
\sum_{i_T,c_T,b,s}P({\cal T}={\rm pass},i_T,c_T,b,s)
I({\cal A};{\cal E}\ |\ i_T,c_T,b,s)
\le 2m\sqrt{\frac{1}{2^{2n}}\sum_b P[(\frac{|C_I|}{n}>p_{a}+\epsilon) \cap
(\frac{|C_T|}{n}\le p_{a})|b]} \ .
\end{equation}
\end{lemm}
{\bf Proof: } \hspace{2mm} The Proof is in Appendix~\ref{exp-bound3}.

We define
$h_b \equiv {P[(\frac{|C_I|}{n}>p_{a}+\epsilon) \cap
(\frac{|C_T|}{n}\le p_{a})|b]} $,
and then: \\
$\sum_{i_T,c_T,b,s}P({\cal T}={\rm pass},i_T,c_T,b,s)
I({\cal A};{\cal E}\ |\ i_T,c_T,b,s)
\le 2m\sqrt{\frac{1}{2^{2n}}\sum_b h_b}$.
The current bound can be dealt with the help of a
random sampling theorem (Hoeffding's law of large numbers\cite{Hoeffding}).
For a long string, the test bits and the
information bits should have similar number of errors
if the test is picked
at random.
The probability that they have
different numbers of errors should go to
zero exponentially fast as shown in the following lemma.
\begin{lemm} \em
For any $\epsilon >0$,\ \  $h_b \le 2 e^{-\frac{1}{2}n\epsilon^2}$.
\label{sec2-lemm}
\end{lemm}
{\bf Proof: } See Appendix~\ref{law-large-num}.

\noindent
As a {\bf Corollary} we get
$\sum_{i_T,c_T,b,s}P({\cal T}={\rm pass},i_T,c_T,b,s)
I({\cal A};{\cal E}\ |\ i_T,c_T,b,s)
\le
2m\sqrt{2 e^{-\frac{1}{2}n\epsilon^2}} = A e^{-\beta n} $,
with $A = 2 m \sqrt 2$ and $\beta = \epsilon^2 / 4$.

Using $I_{Eve}\equiv I({\cal A};{\cal E}|i_T,c_T,b,s)$ and the above
bound we obtain the security criterion 
(see Appendix \ref{app:sec_crit}): 
\begin{equation}
Prob({\rm Test\ Passes\ and\ } I_{Eve} \ge A_{\rm info}
\ e^{-\beta_{\rm info} n} ) \le
 A_{luck} \  e^{-\beta_{luck}n} 
\end{equation}
with
$A_{\rm info} = A_{\rm luck} = \sqrt A$
and $\beta_{\rm info} = \beta_{\rm luck} = \beta/2$.

When choosing $\epsilon = \epsilon_{rel}$ in 
Lemma~\ref{sec2-lemm}, that Lemma also provides the proof that,
once the test passes, there are no more than 
$(p_a + \epsilon_{rel})n$ errors in the information string
(except for exponentially small probability
$2 e^{-\frac{1}{2}n\epsilon^2}$), so that the ECC corrects
these errors.
Thus $A_{rel} = 2$ and $\beta_{rel} = \epsilon_{rel}^2 / 2$, in the
reliability criterion.

The above bound of Eve's information is exponentially small,
but it assumes that ECC codes with 
the desired properties exist.
We present an asymptotic result of security and reliability
using random linear codes (RLC) 
in Appendix~\ref{app:codes-exist} where we analyze RLC 
and we show that such a code provides 
an asymptotic reliability and security, for an allowed error-rate
below 7.56\%.

\section{Summary}
\label{sec:summary}

We proved the security of the Bennett-Brassard (BB84) protocol for
quantum key distribution. Our proof is based on
information-versus-disturbance, on the optimality of symmetric
attacks, on laws of large numbers, and on various techniques that
simplifies the analysis of the problem.

\appendix

\section{Comparison to Other approaches to the security of QKD}
\label{app:sec-proofs} 

    A proof of security of a non practical QKD (based on
the use of quantum computers by the legitimate users) is
provided in~\cite{LC98};
it is based on the assumptions of fault tolerant quantum
computation, hence it can not provide yet a bound on Eve's information.
Apart from this, it is a very strong proof. 
The work of Lo and Chau continues previous works
on quantum privacy amplification~\cite{QPA} and~\cite{Mor96}.

A security claim equivalent to ours (for the same QKD protocol
of~\cite{BB84}) was announced in
Crypto96~\cite{Mayers96} and its details (and some corrections)
are given later on, in publicly announced
drafts~\cite{Mayers98}.
It is obtained using different tools, based on
the security of quantum oblivious transfer
(given bit commitment)~\cite{Yao95,Mayers95}.

Our work provides a rather simple security proof based on the
intuitive concept of information versus disturbance.

Our result does not suffer from any problem related to 
the use of fault-tolerant
computation, which in our case refers to the regular classical
fault-tolerant computation used in classical information
processing steps of any classical protocol:
since we discuss probabilities, the only
effect of a coherent error on the classical computation is to
add some extremely small
probability of error $P(\hbox{\rm coherent classical noise})$
to the final probability of failure (called $P_{luck}$).
On the other hand, doing the same for the quantum coherent noise in
the protocol~\cite{LC98} is not obvious, and still requires a proof.

There are two recent suggestions to attack the security problem
using analogy to quantum error correction~\cite{Shor-99}
and using compression and random sampling~\cite{Ben-Or},
but we have yet to see written drafts of these ideas.

\section{Security of BB84} \label{app:used-bits}

In the paper we prove that used-bits-BB84 is secure.
Let us now present the original BB84 protocol and prove,
by reduction, that its
security follows immediately from the security of the used-bits-BB84
protocol.

The differences between the protocols are only in the first part:

\begin{enumerate}

\item[I.]Creating the sifted key: 

\item

Alice and Bob choose a large integer $n \gg 1$, and a number
$\delta_{\rm num}$, such that
$ 1 \gg \delta_{\rm num} \gg  1/\sqrt{(2 n)}$.
The protocol uses $n'' = (4+\delta_{\rm num})n$ bits.

\item

Alice randomly selects two $n''$-bit strings, $b$ and $i$, 
which are then
used to create qubits:
The string $b$ determines the basis $0\equiv z$, and $1 \equiv x$ of the
qubits.
The string $i$ determines the value (0 or
1) of each of the $n''$ qubits (in the appropriate bases).

\item

Bob randomly selects an $n''$-bit string, $b'$, which determines Bob's
later choice of bases for measuring each of the $n''$ qubits.

\item

Alice generates $n''$ qubits according to her selection of
$b$ and $i$, and sends
them to Bob via a quantum communication channel.

\item
After receiving the qubits, Bob measures in the basis determined by $b'$.

\item
Alice and Bob publish the bases they used; this step should be performed
only after Bob received all the qubits.

\item
All qubits with different bases are discarded by Alice and Bob.
Thus, Alice and Bob finally have $n' \approx n''/2$ bits
for which they used the same bases.
The $n'$-bit string 
would be identical for Alice and Bob
if Eve and natural
noise do not interfere.

\item\label{it:abort}
Alice selects the first $2n$
bits from the $n'$-bit string,  
and the rest of the $n'$ bits are discarded.
If $n'<2n$ the protocol is aborted.

We shall refer to the resulting $2n$-bit string as the sifted key.

\end{enumerate}

The second part of the protocol is identical to the second part of the
used-bits-BB84 protocol.
To prove that BB84 is secure let us modify BB84 by a few steps in a way
that each step can only be helpful to Eve, and the final protocol is the
used-bits-BB84.

Recall that Alice and Bob choose their strings of basis $b$ and $b'$ in
advance. Recall the the two strings are random.
Thus, the first modification below has no influence at all on the
security or the analysis of the BB84 protocol.
Note that after the first modification Alice knows
the un-used bits in advance. 
The second modification is done in a way that Eve can only gain, hence
security of the resulting protocol provides the security of BB84.
The third modification is only ``cosmetic'', 
in order to derive precisely
the used-bits-BB84 protocol. This modification changes nothing in
terms of Eve's ability.

\begin{itemize}

\item 
Let Bob have a quantum memory. 
 Let Alice choose $b'$ instead of Bob at step 3. 
 When Bob receives the qubits at step 5, let
 him keep the qubits in a memory, and tell Alice he received them.
 In step 6, 
let Alice announce $b'$ to Bob, and Bob measure in bases $b'$.

Bob immediately knows which are the used and the un-used bits
(as follows directly from announcing $b$ and $b'$).
Steps 7 and 8 are now combined since Alice and Bob know
all the un-used bits already, and they ignore them,
to be left with $2n$ bits.

\item 
Let Alice generate and send to Bob
only the used bits in step 4, 
and let her ask Eve to send the un-used bits (by telling her
which these are, and also the preparation data for
the relevant subsets, that is---$b_{un-used}$ and 
$i_{un-used}$).
Knowing which are the used bits, and knowing their bases 
$b_{un-used} $ and values $i_{un-used} $
can only help Eve in designing her attack $U'$.

Since Bob never uses the values of the unused bits in the  
protocol (he only ignores them),
he doesn't care if Eve doesn't provide him these bits or provide them to
him without following Alice's preparation request. 

After Bob receives the used and unused bits, let him give Eve the
unused qubits (without measuring them),
and ask her to measure them in bases $b'_{un-used}$.
Having these qubits can only help Eve in designing her optimal final 
measurement.

Since Bob never use the values of the unused bits 
in the rest of the protocol,
he doesn't care if Eve doesn't provide him these values correctly or at
all.

\item

Since Alice and Bob never made any use of the unused bits, 
Eve could have
them as part of her ancilla to start with, 
and Alice could just create $2n$
bits, send them to Bob, and then tell him the bases.

The protocol obtained after this reduction, is a protocol in which 
Eve has
full control on her qubits and on the unused qubits.
Alice and Bob have control on the preparation and measurement of 
the used
bits only.  This is the used-bits BB84, for which we prove 
security in the
text.

\end{itemize}

One important remark is that the exponentially small 
probability that $n' <  2n$ 
in Step~\ref{it:abort} 
(so that the protocol is aborted due to
insufficient number of bits in the sifted key) now becomes a
probability that the reduction fails.

Another important remark is that the issue of high loss rate of qubits
(e.g., due to losses in transmission or detection) can also be handled
via the same reduction.
Thus, our proof applies also to a more practical BB84 protocol 
where high losses are allowed.  

By the way, another practical aspect
is imperfect sources (in which the created states are not
described by a two-level system). This subject
is the issue of recent subtlety
regarding the security of practical schemes, and it is not discussed in
this current work.

\section{Symmetrization}
\label{app:symmetry}

We prove the
optimality of $0/1$ symmetrization.
It is also shown that the symmetrized attacks retain
$0/1$ symmetrization even after the test bits measured.

The fundamental reason symmetrization works, is because
Alice sends Bob bits that have a great deal of symmetry before the key
is defined.
For instance, Alice sends $0$ with probability half, and likewise $1$.
So, Eve can
gain nothing by assuming a particular string was sent 
or by optimizing her
attack for a particular string.
We discuss the symmetrization of the attack $U$,
on $2n$ qubits.
We use $| E_{i_T,i_I,j_T,j_I} \rangle $ for $ |E_{ij}' \rangle $,
when both test and information bits are considered together.

Eve's attack over all the bits is defined:
\begin{equation}
U\ket{0}\ket{i}=\sum_j\ket{E_{i,j}'}\ket{j}
\end{equation}
The following
symmetric attack is defined:
\begin{equation}
\label{eq:symm_eq1}
\ket{E_{i,j}^{'sym}}=\frac{1}{\sqrt{2^{2n}}}
\sum_{m}(-1)^{(i\xor j)\cdot m}
\ket{m}\ket{E_{i\xor m,j\xor m}'}
\end{equation}

This symmetric attack is at least as good as the original attack:
Eve could, as part of her
attack, measure the bits of $\ket{m}$.  
She would then collapse her attack to
$\ket{E_{i\xor m,j\xor m}'}$ uniformly distributed over $m$.  
If the attack $\ket{E_{i,j}'}$
was optimal and if $i,j$ are uniformly distributed, 
then $\ket{E_{i\xor m,j\xor m}'}$ is
also optimal as it only represents a shift in Alice's (already random)
choice of bits.
{}From Eve's perspective the symmetrization is simple.  She
prepares the register: 
$\sum_m\ket{m}$ and uses it as the control bits in
a $Controlled-(\sigma_x\sigma_z)$ (i.e., the controlled-NOT described in
Section~\ref{sec:attack}) onto the bits intended for Bob.  Eve then
applies her unsymmetrized attack $\ket{E_{i,j}'}$.  Following this she
un-applies the bit flips by $Controlled-(\sigma_z\sigma_x)$.

Now we show that information bits are still symmetric
after test bits are measured:
If we first symmetrize the test then the information, we will get the
following:
\begin{equation}
\ket{E_{i_I,i_T,j_I,j_T}^s}=\frac{1}{2^n}\sum_{m_I,m_T}
\ket{m_I}\ket{m_T}(-1)^{(i_I\xor j_I)\cdot m_I}
(-1)^{(i_T\xor j_T)\cdot m_T}
\ket{E_{i_I\xor m_I,i_T\xor m_T,j_I\xor m_I,j_T\xor m_T}}
\end{equation} 
Since $i=i_I i_T$, $j=j_I j_T$, and $m=m_I m_T$, (\ref{eq:symm_eq1}) becomes:
\begin{equation}
\ket{E_{i_I,i_T,j_I,j_T}^s}=\frac{1}{2^n}\sum_{m_I,m_T}
\ket{m_I}\ket{m_T}(-1)^{((i_I i_T)\xor (j_I j_T))\cdot (m_I m_T)}
\ket{E_{i_I\xor m_I i_T\xor m_T,j_I\xor m_I j_T\xor m_T}}
\end{equation} 
These two are identical because:
\begin{eqnarray*}
(-1)^{((i_I i_T)\xor (j_I j_T))\cdot (m_I m_T)}
&=&(-1)^{(i_I\xor j_I)\cdot m_I}
(-1)^{(i_T\xor j_T)\cdot m_T}
\end{eqnarray*}

Hence the bit symmetrization 
is applied independently on test and information.
This is used in the paper when we show that the basis $\ket{\eta_i}$ is
an orthogonal basis.

So now we write the $\ket{E_{i_I,j_I}^s}$ which is the attack once the
test is given:
\begin{equation}
\ket{E_{i_I,j_I}^s}=
\frac{1}{\sqrt{p(j_T|i_I,i_T)}}\ket{E_{i_I,i_T,j_I,j_T}^s}
\end{equation}
where $p(j_T|i_I,i_T)=\sum_{j_I}
\braket{E_{i_I,i_T,j_I,j_T}^s}{E_{i_I,i_T,j_I,j_T}^s}$.  This can be
expanded in the following manner:
\begin{eqnarray*}
p(j_T|i_I,i_T)&=&\sum_{j_I}
\braket{E_{i_I,i_T,j_I,j_T}^s}{E_{i_I,i_T,j_I,j_T}^s}\\
&=&\frac{1}{2^{2n}}\sum_{j_I}\sum_{m_I,m_T}
\braket{E_{i_I\xor m_I,i_T\xor m_T,j_I\xor m_I,j_T\xor m_T}}
{E_{i_I\xor m_I,i_T\xor m_T,j_I\xor m_I,j_T\xor m_T}}\\
&=&\frac{1}{2^{ 2n}}\sum_{c_I}\sum_{m'_I,m_T}
\braket{E_{m'_I,i_T\xor m_T,c_I\xor m'_I,j_T\xor m_T}}
{E_{m'_I,i_T\xor m_T,c_I\xor m'_I,j_T\xor m_T}}\\
\end{eqnarray*}
So in fact, $p(j_T|i_I,i_T)$ is independent of $i_I$ so we may write:
 \begin{equation}
\ket{E_{i_I,j_I}^s}=
\frac{1}{\sqrt{p(j_T|i_T)}}\ket{E_{i_I,i_T,j_I,j_T}^s}
\end{equation}
We can imagine that the information 
bit symmetrization is applied after the test symmetrization, 
and 
after the measurement of the test bits has been done.  
Of course one cannot
apply the information bit symmetrization after 
the information bits have been
measured, because the attack is fixed after all bits have been measured.
\begin{eqnarray*}
\ket{E_{i_I,j_I}^{s_T}}&=&
\frac{1}{\sqrt{p(j_T|i_T)}}\ket{E_{i_I,i_T,j_I,j_T}^{s_T}} \\
&=&\frac{1}{\sqrt{p(j_T|i_T)}}\frac{1}{\sqrt{2^n}}\sum_{m_T}
\ket{m_T}
(-1)^{(i_T\xor j_T)\cdot m_T}
\ket{E_{i_I,i_T\xor m_T,j_I,j_T\xor m_T}}
\end{eqnarray*}
Now, we can symmetrize information bits separately
\begin{equation}
\ket{E_{i_I,j_I}^s}=\frac{1}{\sqrt{2^n}}\sum_{m_I}
\ket{m_I}
(-1)^{(i_I\xor j_I)\cdot m_I}
\ket{E_{i_I\xor m_I,j_I\xor m_I}^{s_T}}
\ .
\end{equation}

The above equation is used in Appendix \ref{app:eta-basis}
in that the attack still has $0/1$ symmetry after the test
bits are announced.

{}From Alice and Bob's prospective, Eve's attack 
has been averaged over all possible inputs.
The probability of a given error is now independent of $i$:
\begin{eqnarray*}
P(\ket{j}=\ket{i\xor c})_{sym}&=&
\braket{E_{i,i\xor c}^{s}}{E_{i,i\xor c}^{s}} \\
&=&\frac{1}{2^{2n}}\sum_{m,m'}
\braket{E_{i\xor m,i\xor m \xor c}}{E_{i\xor m',i\xor m' \xor c}}
\braket{m}{m'}(-1)^{c\cdot(m\xor m')} \\
&=&\frac{1}{2^{2n}}
\sum_m\braket{E_{i\xor m,i\xor m \xor c}}{E_{i\xor m,i\xor m \xor c}}\\
&=&
\frac{1}{2^{2n}}\sum_{i'}\braket{E_{i',i' \xor c}}{E_{i',i' \xor c}} \\
&=&\frac{1}{2^{2n}}\sum_{i'}P(\ket{j}=\ket{i'\xor c})_{unsym}
\end{eqnarray*}
In the second to last step we change variables $i'=m\xor i$.

The symmetrization makes the symmetrized error distribution equal to the
averaged error distribution on the original attack.  Hence the
average error rate for Alice and Bob is the same for the two attacks.
This is important as we assume that Eve losses nothing by applying
a symmetric attack.

We have calculated the fact that the error does not depend on
the sent bits for the information+test string.
Similarly, the probabilities $c_T$ and $c_I$ are independent of
Alice's choice of bits.
This can be shown easily by modifying the above calculations
in very straightforward ways to consider any choice of substring.
Intuitively it is clear that if the whole error string
has a probability distribution
independent of $i$, then any randomly selected subsets would also.

\section{A Proof for Lemma \ref{basis-eta}}
\label{app:eta-basis}

First we show that $\braket{\phi_l}{\phi_{l\xor k}}$ does not
depend on $l$ for a symmetrized attack:

From Appendix \ref{app:symmetry} we know that the
symmetric attack after measurement can be written as: 
\[\ket{E_{i,j}}=
\frac{1}{\sqrt{2^n}}\sum_m(-1)^{(i\xor j)\cdot m}
\ket{m}\ket{E^n_{i\xor m,j\xor m}} \]
with $\ket{E^n_{i,j}}$ being the attack symmetrized only over test bits,
after the Bob's test bits have been announced.

Now, we may compute $\braket{\phi_{l}}{\phi_{l\xor k}}$
for this (symmetric) attack:
\begin{eqnarray*}
\braket{\phi_{l}}{\phi_{l\xor k}}&=&\sum_j
\braket{E_{l,j}}{E_{l\xor k,j\xor k}}\\
&=&\frac{1}{2^{n}}\sum_{j,m}
\braket{E^n_{l\xor m,j\xor m}}{E^n_{l\xor m \xor k,j\xor m \xor k}}\\
\txtline{Change variables $m'=m\xor l$ and $j'=j\xor l$, so $j\xor m=j'\xor m'$}
\braket{\phi_{l}}{\phi_{l\xor k}}&=&
\frac{1}{2^{n}}\sum_{j',m'}
\braket{E^n_{m',j'\xor m'}}{E^n_{m' \xor k,j'\xor m' \xor k}}\\
\end{eqnarray*}
Therefore, $\braket{\phi_{l}}{\phi_{l\xor k}}$ does not depend on $l$
for the $0/1$-symmetrized attacks.
Thus, we can define $\braket{\phi_{l}}{\phi_{l\xor k}}=\Phi_k$.

Now, by definition of $\eta$,
we have $\braket{\eta_{i}}{\eta_{j}}=\frac{1}{2^{2n}}\sum_{l,m}(-1)^{l\cdot i}
(-1)^{j\cdot m}\braket{\phi_{l}}{\phi_{m}}$.
Setting $k = l \xor m$
\begin{eqnarray*}
\braket{\eta_{i}}{\eta_{j}}&=&\frac{1}{2^{2n}}\sum_{l,k}(-1)^{l\cdot i}
(-1)^{j\cdot (l\xor k)}\braket{\phi_{l}}{\phi_{l\xor k}}\\
&=&\frac{1}{2^{n}}\sum_{k}(-1)^{j\cdot k}\Phi_k
\left(\frac{1}{2^n}\sum_{l}(-1)^{l\cdot (i\xor j)}\right)\\
&=&\frac{1}{2^{n}}\sum_{k}(-1)^{j\cdot k}\Phi_k \ \delta_{i,j}
\end{eqnarray*}

When $i \ne j$, $\delta_{i,j}$ is zero, and thus
\< \braket{\eta_{i}}{\eta_{j}}=0 .\>
\hfill {\bf QED}

\section{A Proof for Lemma \ref{d-sqr-is-p-err}}
\label{app:d-sqrt-is}

First note that the norm of $\eta_{i}$ satisfies
\begin{eqnarray} \label{norm-eta}
d_i^2 = \braket{\eta_{i}}{\eta_{i}}&=& \frac{1}{2^{2n}}
	\sum_{l}   \sum_{l'} (-1)^{(l\xor l')\cdot i}
	\braket{\phi_{l}}{\phi_{l'}}=   \nonumber \\
	&=&\frac{1}{2^{2n}}\sum_{l}   \sum_{k} (-1)^{i\cdot k}
	\braket{\phi_{l}}{\phi_{l\xor k}}   \nonumber \\
        &=& \frac{1}{2^{2n}}\sum_{l,k,j}
	(-1)^{i\cdot k}
	\braket{E_{l,j}}{E_{l\xor k,j \xor k}}
\end{eqnarray}
which was calculated using
\begin{eqnarray*}
\braket{\phi_{l}}{\phi_{l\xor k}}&=&\sum_{j}\sum_{j'}
	\braket{E_{l,j}}{E_{l\xor k,j'}}
	\braket{l\xor j}{l\xor k \xor j'}\\
	&=&\sum_{j}
	\braket{E_{l,j}}{E_{l\xor k,j \xor k}}\\
\end{eqnarray*}

After the test and the unused bits are announced we have:
\[\ket{i}\rightarrow\sum_j\ket{E_{i,j}}\ket{j} \]
The basis transformation is
\[\ket{k}^o=\frac{1}{\sqrt{2^n}}\sum_i (-1)^{i\cdot k}\ket{i} \]
Once Eve's attack is given in one basis it can be
calculated in the other basis (due to linearity) so that:
\[\ket{i}^o\rightarrow\sum_j\ket{E_{i,j}^o}\ket{j}^o \]
with
\[\ket{E_{k,l}^o}=\frac{1}{2^n}
\sum_{i,j}(-1)^{i \cdot k}(-1)^{j \cdot l}
\ket{E_{i,j}}\]

For any given outcomes of the test and unused bits, the probability
$P(c_I^o \ | \ i_T,j_T,b,s)$ of an
error string $c$ over the information bits in the opposite basis is
\begin{eqnarray*}
P(c_I^o \ | \ i_T,j_T,b,s) &=&
	\frac{1}{2^n}\sum_{k}\braket{E_{k,k\xor c}^o}{
	E_{k,k\xor c}^o}=\\
 &=& \frac{1}{2^n}\sum_{k} \sum_{i,j} \sum_{i',j'}\frac{1}{2^{2n}}
(-1)^{(i\xor i')\cdot k} (-1)^{(j\xor j')\cdot (k\xor c)}
\braket{E_{i,j}}{E_{i',j'}}\\
 &=& \frac{1}{2^{2n}}\sum_{i,j} \sum_{i',j'}
	\left(\frac{1}{2^n}\sum_{k} (-1)^{k\cdot (i\xor i' \xor j\xor j')}
	\right)
	(-1)^{c\cdot (j\xor j')}
	\braket{E_{i,j}}{E_{i',j'}}\\
\txtline{The sum over $k$ is non zero only when
	$i\xor i' = j \xor j' \stackrel{\Delta}{=} h$}
 &=& \frac{1}{2^{2n}}\sum_{i,j,h} (-1)^{c\cdot h}
	\braket{E_{i,j}}{E_{i\xor h,j\xor h}}\\
&=&\braket{\eta_{c}}{\eta_{c}} = d_c^2
\end{eqnarray*}
where the last equalities are due to the calculation
of the norm of $\eta$
in Eq.~(\ref{norm-eta}).
\hfill {\bf QED}

\section{The Error Correction Code}
\label{app:ECC}

After all the quantum bits are sent and Bob has made all his 
measurements, Alice announces an $(n,k,d)$ ECC, 
${\cal C}$, and the corresponding syndrome of her information bit
string, $i_I$.  In our protocol, ${\cal C}$, is announced by sending 
the $r=(n-k)$ rows $v_j, j=1, \ldots, r$ of a
maximum rank $r \times n$ 
generator matrix, ${\cal H}$, of its dual code, 
$S_{\mathbf s} = 
{\cal C}^{\perp} = \{v_s \in \{0,1\}^n \ | \ v_s \cdot c = 0
\ \forall c \in {\cal C}\}$ [and $\{v_1, \ldots, v_r\}$ is a basis of
$S_{\mathbf s} = {\cal C}^{\perp}$].
The code-syndrome
sent by Alice comprises $\xi^{Alice}\equiv {\cal H}\cdot i_I$.  
Bob can now
calculate $\xi^{Bob}\equiv {\cal H}\cdot j_I$.  Since $j_I=i_I\xor c_I$,
he can learn ${\cal H}\cdot c_I=\xi^{Alice}\xor\xi^{Bob}$.  Hence, Bob
has the syndrome of his error string and can correct the errors if
$|c_I| \leq (d-1)/2$. 

In the case of a one bit key, 
the final key is extracted as $b = i_I\cdot v$, 
where $v\in \{0,1\}^n$
is the string defined for PA. 
Of course $v$ is chosen such that $v \notin
S_{\mathbf s}$.
Before we proceed, 
we shall choose once and for all a subspace $S_{\mathbf s}^c$ of 
$\{0,1\}^n$ containing $v$ and such that 
$S_{\mathbf s}$ and $S_{\mathbf s}^c$ are complementary; this means
that $S_{\mathbf s}^c$ is an
$(n-r)$ dimensional subspace,
such that $ S_{\mathbf s}^c \xor S_{\mathbf s}= \{0,1\}^n$, and 
$ S_{\mathbf s}^c \cap S_{\mathbf s} =\{0\}$. 
Of course $S_{\mathbf s}^c \neq S_{\mathbf s}^{\perp} = {\cal C}$.
Note that $S_{\mathbf s}^c$ is not unique (but is easy to construct
after augmenting the set of vectors $\{v_1, \ldots, v_r, v\}$ to a
basis of $\{0,1\}^n$).

The preceding error correction procedure, however, provides additional
information to Eve. 
In particular, since Eve knows ${\cal H}$ (i.e., the generator matrix
of $S_{\mathbf s}$) 
and the syndrome $\xi^{Alice}\equiv {\cal H}\cdot i_I$, 
Eve knows which coset 
${\cal C}_{\xi} = \{i_I \in \{0,1\}^n \ | \ {\cal H} i_I = \xi\}$
of $\cal C$ Alice's information bit string, $i_I$, belongs to. 
That is,
she learns it belongs to ${\cal C}_{\xi}$ for $\xi = \xi^{Alice}$. 

Let $i_\xi$ denote a code word in 
the coset ${\cal C}_{\xi}$ (we dropped the index ${}_I$, for
convenience).
Note
that for any $i_\xi \in {\cal C}_{\xi}$ we have
$${\cal C}_\xi= \{(i_\xi \oplus c \ | \ c \in {\cal C}\},$$
that is, all other code words in 
the coset ${\cal C}_{\xi}$ can be obtained from one code word
in ${\cal C}_{\xi}$ by calculating its exclusive-or with all the code
words in the original linear code ${\cal C}$.
We now choose arbitrarily an $i_\xi$ in each ${\cal C}_\xi$. Those
arbitrary\footnote{In fact there is a unique
$i_{\xi} \in (S_{\mathbf s}^c)^{\perp} \cap {\cal C}_\xi$ 
but that is irrelevant
for our proofs.}
but fixed strings will often be referred to in the next appendix.

\section{Eve's Information Versus the disturbance}
\label{app:SD-vs-d}

In this appendix we do not prove Lemma~\ref{sd-lemm} immediately.
We prove it later on, in the second
subsection (the tight bound). 
For simplicity of the presentation, we first prove
another Lemma which leads to a loose bound (with an additional 
factor of $2^r$), for which the derivation is simpler.
The bulk of the loose bound was derived in~\cite{BBBGM}, and the tight
bound is an improvement over that derivation.
The loose bound lead to a much worse threshold for $p_{allowed}$
(less than 1\%, instead of 7.56\% derived from the tight bound), 
and this is the motivation for deriving the tight bound.
One can skip directly to
the second subsection if desired.

Both the loose and the tight bound are derived using   
the fact that
the Shannon distinguishability between the parity 0 density matrix,
$\rho_0$, and the parity 1 density matrix, $\rho_1$,  
is bounded by the trace norm of $\rho_0-\rho_1$, and using the fact 
that  
the one can easily calculate this trace-norm when the purified states
are given by Eq.~\ref{eq:phi}. 

\subsection{The Loose Bound (BBBGM)}
\label{appen-BBBGM}

We have already defined a purification of Eve's state:
$\ket{\phi_{i_I}} = \sum_l (-1)^{i_I\cdot l} \ket{\eta_l}$
The density matrix for such a $\ket{\phi_{i_I}}$ is
\begin{equation}
\rho^{i_I}=\ket{\phi_{i_I}}\bra{\phi_{i_I}}
	=\sum_{l,l'}
		(-1)^{i_I(l\xor l')} d_l d_{l'}
		\ket{\hat\eta_l}\bra{\hat\eta_{l'}}
\end{equation}
Recall that the final key is computed as $v\cdot i_I$. 
Eve does not know
$i_I$, but she knows from 
the announced syndrome 
that $i_I$ is in the coset 
${\cal C}_\xi$ for $\xi \equiv \xi^{Alice}$. Hence, in order to know
the key, Eve must distinguish between the states 
$i_I = i_\xi \oplus c$ in 
${\cal C}_\xi$ that 
give parity zero and the states 
$i_I = i_\xi \oplus c$ in 
${\cal C}_\xi$ that 
give parity one. 
For $b\in \{0,1\}$ the reduced density
matrix is
\begin{eqnarray*}
\rho_b&=&\frac{1}{2^{n-(r+1)}}
\sum_{\stackrel{c \in {\cal C}}{v(i_\xi \oplus c)=b}} 
\rho^{i_\xi \oplus c}=\\
	&=&\frac{1}{2^{n-(r+1)}}
\sum_{\stackrel{c \in {\cal C}}{v(i_\xi \oplus c)=b}}
		\sum_{l,l'}
		(-1)^{(i_\xi \oplus c)(l\xor l')} d_l d_{l'}
		\ket{\hat\eta_l}\bra{\hat\eta_{l'}}
\end{eqnarray*}
where the sum 
is over $c$ that satisfy both the condition of being a
code word, and the condition
of leading to the particular parity $b$ for
the PA.

{\bf Lemma.}
Let ${\cal C}$ be any linear code in $\{0, 1\}^n$ and $a \in \{0,1\}^n$ be 
such that $a \notin {\cal C}^{\perp}$ then
\begin{equation} \label{zerosum}
\sum_{c \in {\cal C}} (-1)^{c \cdot a} = 0
\end{equation}

\ \newline
{\bf Proof}.---
Let $\{w_1, \ldots, w_k\}$ be a basis of ${\cal C}$. Define $t \in \{0,1\}^k$
by $t_\alpha = w_\alpha \cdot a, \ 1 \leq \alpha \leq k$; 
$a \notin {\cal C}^{\perp}$ means that $t$ is not the zero string.
Let now 
$h: \{0,1\}^k \rightarrow {\cal C}$ be defined by 
$h(s) = \sum_{1 \leq \alpha \leq k} s_{\alpha}w_{\alpha}$; then
$h(s) \cdot a = \sum s_{\alpha} w_{\alpha} \cdot a = \sum s_{\alpha} t_\alpha 
= s \cdot t$ and so
$$\sum_{c \in {\cal C}} (-1)^{c \cdot a} =
  \sum_{s} (-1)^{h(s) \cdot a} =
  \sum_{s} (-1)^{s \cdot t} = 0$$

\ \newline
{\bf Lemma \label{sd-lemm-bbbgm}}
The Shannon distinguishability 
between the parity 0 and the parity 1 of the
information bits over any PA string, $v$, 
is bounded above by the following inequality:
\begin{equation}
SD_v\le 2^{r}\left[\alpha+\frac{1}{\alpha}
\sum_{|l|\ge\frac{\hat v}{2}}d_l^2\ \right] ,
\end{equation}
where $\hat v$ is the  minimum 
weight of $v\xor v_s$ for any $v_s \in S_{\mathbf s}$, and
$\alpha$ is any positive constant.

\ \newline

{\bf Proof}.---
The Shannon distinguishability between the parity 0 and the parity 1
is bounded by the trace norm of $\rho_0-\rho_1$ (\cite{BBBGM,FG}).
Let us calculate the required bound:
\begin{eqnarray*}
\rho_0-\rho_1&=&\frac{1}{2^{n-(r+1)}}
\sum_{c \in {\cal C}} (-1)^{(i_\xi \oplus c)v}
		\sum_{l,l' }
		(-1)^{(i_\xi \oplus c)(l\xor l')} d_l d_{l'}
		\ket{\hat\eta_l}\bra{\hat\eta_{l'}}\\
	&=&\frac{1}{2^{n-(r+1)}} \sum_{l,l'}
		\left(\sum_{c \in {\cal C}}
			(-1)^{(i_\xi \oplus c)(l\xor l'\xor v)}\right)
		d_l d_{l'} \ket{\hat\eta_l}\bra{\hat\eta_{l'}}\\
	&=&\frac{1}{2^{n-(r+1)}} \sum_{l,l'}
               (-1)^{i_\xi (l \oplus l' \oplus v)}
		\left(\sum_{c \in {\cal C}}
			(-1)^{c(l\xor l'\xor v)}\right)
		d_l d_{l'} \ket{\hat\eta_l}\bra{\hat\eta_{l'}}\\
\end{eqnarray*}
{}From equation (\ref{zerosum}) we know the sum over ${\cal C}$ is zero
except when $l \oplus l' \oplus v \in {\cal C}^{\perp} = S_{\mathbf s}$,
i.e. when $l' = l \oplus v \oplus v_{\mathbf s}$ for 
some $v_{\mathbf s} \in S_{\mathbf s}$. As a consequence:
\begin{eqnarray*}
\rho_0-\rho_1&=& 2 \sum_{v_s \in S_{\mathbf s} } (-1)^{i_\xi \cdot v_s}
  \sum_l d_l d_{l\xor v \xor v_s}
		\ket{\hat\eta_l}\bra{\hat\eta_{l\xor v \xor v_s}}
\end{eqnarray*}
The trace norm of this matrix serves as a bound on the information Eve
receives.
\begin{eqnarray*}
SD_v&\le& \frac{1}{2}Tr|\rho_0-\rho_1| 
\end{eqnarray*}

Using the above and making use of the triangle inequality for
the Trace norm, the following is obtained: 

\begin{eqnarray*}
SD_v&\le& 
Tr \big| \sum_{v_s \in S_{\mathbf s}} (-1)^{i_\xi \cdot v_s}
  \sum_l d_l d_{l\xor v \xor v_s}
		\ket{\hat\eta_m}\bra{\hat\eta_{m\xor v \xor v_s}} \ \big| \\
&=&\frac{1}{2}Tr| \sum_{v_s \in S_{\mathbf s}} (-1)^{i_\xi \cdot v_s}
 \sum_{l}
		d_l d_{l\xor v \xor v_s}
		\left(\ket{\hat\eta_l}\bra{\hat\eta_{l\xor v \xor v_s}}
+ \ket{\hat\eta_{l\xor v \xor v_s}}\bra{\hat\eta_l}\right) | \\ 
&\le& \sum_{v_s \in S_{\mathbf s}}
\sum_l d_l d_{l\xor v \xor v_s}(\ \frac{1}{2}Tr|\ (
\ket{\hat\eta_l}\bra{\hat\eta_{l\xor v \xor v_s}}
+ \ket{\hat\eta_{l\xor v \xor v_s}}\bra{\hat\eta_l})\ |\ ) \\
&=&\sum_{v_s \in S_{\mathbf s} }
\sum_l d_l d_{l\xor v \xor v_s} 
\end{eqnarray*}
Now we will concern ourselves with bounding each of the terms
$\sum_l d_l d_{l\xor w_s}$,
where $w_s = v \xor v_s$. 
\begin{eqnarray*}
\sum_l d_l d_{l\xor w_s}&=&
	\sum_{|l|>\frac{|w_s|}{2}} d_l d_{l\xor w_s}
	+ \sum_{|l|\le\frac{|w_s|}{2}} d_l d_{l\xor w_s} \\
&=&\sum_{|l|>\frac{|w_s|}{2}} d_l d_{l\xor w_s}
	+ \sum_{|l'\xor w_s|\le\frac{|w_s|}{2}} d_{l'\xor w_s} d_{l'} \\
\end{eqnarray*}
If $|l' \oplus w_s| \leq \frac{|w_s|}{2}$ then 
$  |w_s| = 
             |l' \oplus w_s \oplus l'| 
        \leq |l' \oplus w_s| + |l'| 
        \leq \frac{|w_s|}{2} + |l'|$ 
and so $|l'| \geq \frac{|w_s|}{2}$.
Therefore, 
\begin{eqnarray*}
\sum_{|l|>\frac{|w_s|}{2}} d_l d_{l\xor w_s}
	+ \sum_{|l'\xor w_s|\le\frac{|w_s|}{2}} d_{l'\xor w_s} d_{l'}
&\le&\sum_{|l|\ge\frac{|w_s|}{2}} d_l d_{l\xor w_s}
	+ \sum_{|l'|\ge\frac{|w_s|}{2}} d_{l'\xor w_s} d_{l'} \\
&=&2\sum_{|l|\ge\frac{|w_s|}{2}} d_l d_{l\xor w_s} \\
&=&\frac{1}{\alpha}\sum_{|l|\ge\frac{|w_s|}{2}}2 d_l 
(\alpha d_{l\xor w_s}) \\
&\le&\frac{1}{\alpha} \sum_{|l|\ge\frac{|w_s|}{2}} [d_l^2
 + \alpha^2 d_{l\xor w_s}^2] \\
&=&\alpha\sum_{|l|\ge\frac{|w_s|}{2}} d_{l\xor w_s}^2 
+ \frac{1}{\alpha}\sum_{|l|\ge\frac{|w_s|}{2}}d_l^2 \\
\end{eqnarray*}
where the last three steps are true 
for any real $\alpha$, and real $d_l,d_{l\xor w_s}$.

Due to the fact that the $d_l^2$ form a probability distribution,
any sum of them is less than or equal to unity.
\begin{eqnarray*}
\sum_l d_l d_{l\xor w_s}
&\le&\alpha+\frac{1}{\alpha}\sum_{|l|\ge\frac{|w_s|}{2}}d_l^2
\\
&\le&\alpha+\frac{1}{\alpha}\sum_{|l|\ge\frac{\hat{v}}{2}}d_l^2
\end{eqnarray*}
%
where $\hat{v} = \min_{v_s} | v \xor v_s |$ 
(remember that $w_s = v \oplus v_s$). Summing over all $v_s \in S_{\mathbf s}$
now leaves:
\begin{equation}
SD_v\le 2^{r}
\left[\alpha+\frac{1}{\alpha}\sum_{|l|\ge\frac{\hat{v}}{2}}d_l^2\right]
\end{equation}
\mbox{ } \hfill {\bf QED}

The BBBGM result gives an upper bound 
for Eve's information about the bit
defined by this privacy amplification string $v$.
To prove security in case of $m$ bits in the final key,
we prove security of each bit as follows:
for each bit in the key we assume that 
Eve is given the ECC information and in addition, she is also
given all the other
bits in the key.  This is like using a code with more
parity check strings $2^{r+m-1}$ (or less code words), hence the
previous result holds with

\begin{equation}
SD_v\le 2^{r+m-1} \left[   \alpha +
          \frac{1}{\alpha}\sum_{|l|\ge\frac{\hat v}{2}}d_l^2\ \right]
\  .
\end{equation}

Following the proof of the above Lemma, one can see that it is not a
tight bound since we sum over $2^r$ terms while most
of them are much smaller than the term (terms) with the minimal
$\hat v$.

\subsection{Eve's Information on one bit -- Tight Bound}
\label{tight-bou}

We now show an improved technique, by defining a basis
for the purification of the code words (instead of a basis for
all the purification).

We will now make a finer analysis of Eve's state after she learns
the parity matrix and the syndrome $\xi = \xi^{Alice}$.
We start again from the equality:
\begin{equation}
\ket{\phi_{i_I}} = \sum_l (-1)^{i_I\cdot l} \ket{\eta_l}
\end{equation}
First, any $l \in \{0,1\}^n$ has a unique
representation $l = m \oplus n$ with $m \in S_{\mathbf s}^c$ and
$n \in S_{\mathbf s}$. 
Next, for any $i_I \in {\cal C}_\xi$ we have $i_I = i_\xi \oplus c$ for
some $c \in {\cal C}$ and thus for any $n \in S_{\mathbf s}$ we get
$i_I \cdot n = (i_\xi \oplus c) \cdot n = i_\xi \cdot n$ [because
$n \in S_{\mathbf s} ={\cal C}^\perp$]. 
Putting those two remarks together we get:

\begin{eqnarray*}
\ket{\phi_{i_I}}&=& \sum_{m\in S_{\mathbf s}^c}\sum_{n\in S_{\mathbf s}}
 (-1)^{i_I\cdot (m\xor n)} \ket{\eta_{m\xor n}}\\
&=&\sum_{m\in S_{\mathbf s}^c}(-1)^{i_I\cdot m}\sum_{n\in S_{\mathbf s}}
 (-1)^{i_I\cdot n} \ket{\eta_{m\xor n}} \\
&=&\sum_{m\in S_{\mathbf s}^c}(-1)^{i_I\cdot m}\sum_{n\in S_{\mathbf s}}
 (-1)^{i_\xi \cdot n} \ket{\eta_{m\xor n}} \\
&=&\sum_{m\in S_{\mathbf s}^c}
		(-1)^{i_Im} \ket{\eta'_m}
\end{eqnarray*}
where $\eta'_m$ is of course defined for each $m \in S_{\mathbf s}$ by
\begin{equation}
\ket{\eta'_m} = \sum_{n\in S_{\mathbf s}}
(-1)^{i_\xi \cdot n} \ket{\eta_{m\xor n}}
\end{equation}
Now, since $\braket{\eta_{m_1 \oplus n_1}}{\eta_{m_2 \oplus n_2}}=0$ except
when $m_1 \oplus n_1 = m_2 \oplus n_2$, which implies $m_1 = m_2$, 
the $\eta_m$'s are orthogonal. If $d'_m$ is the length of 
$\eta'_m$, we can then write
\[ \eta'_m = d'_m \hat\eta'_m \]
with the $\hat\eta'_m$'s normalized and orthogonal and
\[ d'{}^2_m = \sum_{n\in S_{\mathbf s}} d_{m\xor n}^2 \]
and the density matrix 
for $\ket{\phi_{i_I}}$ reduces to:
\begin{eqnarray*}
\rho^{i_I}&=&\ket{\phi_{i_I}}\bra{\phi_{i_I}}=\\
	&=&\sum_{m,m'\in S^c_{\mathbf s}}
		(-1)^{i_I(m\xor m')} d'_m d'_{m'}
		\ket{\hat\eta'_m}\bra{\hat\eta'_{m'}}
\end{eqnarray*}
Recall that the final key is computed as $b = v\cdot i_I$. 
Of course, Eve does not
know
$i_I$, but she knows from the announced syndrome $\xi = \xi^{Alice}$ 
that $i_I \in {\cal C}_\xi = \{ i_\xi \oplus c \ | \ c \in {\cal C} \}$
and wants to determine $b$.
For $b\in \{0,1\}$ the reduced density
matrix is
\begin{eqnarray*}
\rho_b&=&\frac{1}{2^{n-(r+1)}}
\sum_{\stackrel{c \in {\cal C}}{(i_\xi \oplus c)v=b}} \rho^{i_\xi \oplus c}=\\
	&=&\frac{1}{2^{n-(r+1)}}
\sum_{\stackrel{c \in {\cal C}}{(i_\xi \oplus c)v=b}}
		\sum_{m,m'\in S^c_{\mathbf s}}
		(-1)^{(i_\xi \oplus c)(m\xor m')} d'_m d'_{m'}
		\ket{\hat\eta'_m}\bra{\hat\eta'_{m'}}
\end{eqnarray*}
\begin{lemm} \em \ref{sd-lemm}
The Shannon distinguishability 
between the parity 0 and the parity 1 of the
information bits over any PA string, $v$, 
is bounded above by the following inequality:
\begin{equation}
SD_v\le\alpha+\frac{1}{\alpha}\sum_{|l|\ge\frac{\hat v}{2}}d_l^2\ ,
\end{equation}
where $\hat v$ is the  minimum weight of $v\xor v_s$ for any $v_s \in S_{\mathbf s}$, and
$\alpha$ is any positive constant.
\end{lemm}

\ \newline
\ {\bf Proof:}
The Shannon distinguishability between the parity 0 and the parity 1
is bounded by the trace norm of $\rho_0-\rho_1$:
\begin{eqnarray*}
\rho_0-\rho_1&=&\frac{1}{2^{n-(r+1)}}
\sum_{c \in {\cal C}} (-1)^{(i_\xi \oplus c)v}
		\sum_{m,m'\in S^c_{\mathbf s}}
		(-1)^{(i_\xi \oplus c)(m\xor m')} d'_m d'_{m'}
		\ket{\hat\eta'_m}\bra{\hat\eta'_{m'}}\\
	&=&\frac{1}{2^{n-(r+1)}} \sum_{m,m'\in S^c_{\mathbf s}}
		\left(\sum_{c \in {\cal C}}
			(-1)^{(i_\xi \oplus c)(m\xor m'\xor v)}\right)
		d'_m d'_{m'} \ket{\hat\eta'_m}\bra{\hat\eta'_{m'}}\\
	&=&\frac{1}{2^{n-(r+1)}} \sum_{m,m'\in S^c_{\mathbf s}}
	        (-1)^{i_\xi (m\xor m'\xor v)}
		\left(\sum_{c \in {\cal C}}
			(-1)^{c \cdot (m\xor m'\xor v)}\right)
		d'_m d'_{m'} \ket{\hat\eta'_m}\bra{\hat\eta'_{m'}}\\
\end{eqnarray*}
Applying equality (\ref{zerosum}) the sum indexed by $c$ is zero except
when $m \oplus m' \oplus v \in {\cal C}^\perp = S_{\mathbf s}$. But
$m \oplus m' \oplus v \in S_{\mathbf s}^c$ because $m, m' \ {\rm and} \ 
v \in S_{\mathbf s}^c$.
This implies $m \oplus m' \oplus v \in S_{\mathbf s} \cap S_{\mathbf s}^c =
\{0\}$ and thus $m' = m \oplus v$.
Of course, with $m \oplus m' \oplus v = 0$, the sum indexed by $c$ is 
$2^k = 2^{n-r}$ and the coefficient $(-1)^{i_\xi (m \oplus m' \oplus v)}$
is 1. Therefore $\rho_0 - \rho_1$ takes the very simple form:
\begin{eqnarray*}
\rho_0-\rho_1&=& 2 \sum_{m\in S^c_{\mathbf s}}
		d'_m d'_{m\xor v}
		\ket{\hat\eta'_m}\bra{\hat\eta'_{m\xor v}}
\end{eqnarray*}
As usual, the trace norm of this matrix serves as a bound on the information Eve
receives.
It is
\begin{eqnarray*}
SD_v&\le& \frac{1}{2}Tr|\rho_0-\rho_1| 
\end{eqnarray*}

First
note that $v$ is in $S^c_{\mathbf s}$ and $S^c_{\mathbf s}$ is closed
under addition.  Further the set $S^c_{\mathbf s}$ is the same as 
$v\xor S^c_{\mathbf s}$.  Then the set defined by $m\in S^c_{\mathbf s}$
is identical to the set $m\xor v\in S^c_{\mathbf s}$.  
We will use this identity to obtain the following
inequality:

\begin{eqnarray*}
SD_v&\le& Tr| \sum_{m\in S^c_{\mathbf s}}
		d'_m d'_{m\xor v}
		\ket{\hat\eta'_m}\bra{\hat\eta'_{m\xor v}}| \\
&=&\frac{1}{2}Tr| \sum_{m\in S^c_{\mathbf s}}
		d'_m d'_{m\xor v}
		\ket{\hat\eta'_m}\bra{\hat\eta'_{m\xor v}}
+  \sum_{m\xor v\in S^c_{\mathbf s}}
		d'_m d'_{m\xor v}
		\ket{\hat\eta'_m}\bra{\hat\eta'_{m\xor v}} | \\
&=&\frac{1}{2}Tr| \sum_{m\in S^c_{\mathbf s}}
		d'_m d'_{m\xor v}
		\ket{\hat\eta'_m}\bra{\hat\eta'_{m\xor v}}
+  \sum_{m\in S^c_{\mathbf s}}
		d'_{m\xor v} d'_m
		\ket{\hat\eta'_{m\xor v}}\bra{\hat\eta'_m} | \\
&=&\frac{1}{2}Tr| \sum_{m\in S^c_{\mathbf s}}
		d'_m d'_{m\xor v}
		(\ket{\hat\eta'_m}\bra{\hat\eta'_{m\xor v}}
+  \ket{\hat\eta'_{m\xor v}}\bra{\hat\eta'_m}) | \\
&\le&\frac{1}{2}\sum_{m\in S^c_{\mathbf s}} d'_m d'_{m\xor v}
  Tr|\ket{\hat\eta'_m}\bra{\hat\eta'_{m\xor v}}+
  \ket{\hat\eta'_{m\xor v}}\bra{\hat\eta'_m}| \\
&=&\sum_{m\in S^c_{\mathbf s}} d'_m d'_{m\xor v} \\
\end{eqnarray*}
Now we wish to give a bound in terms of the original $d$'s.  
Let us define 
\begin{eqnarray*}
\Gamma_{\hat v}&=&\{m\in S^c_{\mathbf s} \mid |m\xor n|\ge\hat v/2
\ \forall n \in S_{\mathbf s}\}
\end{eqnarray*}
where $\hat{v}$ was defined in the statement of the lemma. We claim
that for any $m \in S_{\mathbf s}^c$, 
either $m\in \Gamma_{\hat v}$ or $m\xor v\in \Gamma_{\hat v}$. Indeed,
if it were not so, there would be 
$n_1 \in S_{\mathbf s}$ and 
$n_2 \in S_{\mathbf s}$ such that $|m \oplus n_1| < \hat{v}/2$ and
$|m \oplus v \oplus n_2| < \hat{v}/2$. But then
$|n_1 \oplus n_2 \oplus v| =
 |m \oplus n_1 \oplus m \oplus n_2 \oplus v| <
 \hat{v}/2 + \hat{v}/2$ which,
since $n_1 \oplus n_2 \in S_{\mathbf s}$, 
contradicts the definition of $\hat{v}$.

We now use the claim to break up the sum bounding $SD_v$ and prove the lemma.
\begin{eqnarray*}
SD_v&\le&\sum_{m\in S^c_{\mathbf s}} d'_m d'_{m\xor v} \\
&\le&\left(\sum_{\stackrel{m\in S^c_{\mathbf s}}
{m\in\Gamma_{\hat v}}} d'_m d'_{m\xor v}
+ \sum_{\stackrel{m\in S^c_{\mathbf s}}
{m\xor v\in\Gamma_{\hat v}}} d'_m d'_{m\xor v}\right) \\
&=&\left(\sum_{\stackrel{m\in S^c_{\mathbf s}}
{m\in\Gamma_{\hat v}}} d'_m d'_{m\xor v}
+ \sum_{\stackrel{m\xor v\in S^c_{\mathbf s}}
{m\in\Gamma_{\hat v}}} d'_m d'_{m\xor v}\right) \\
&=&2\sum_{m\in\Gamma_{\hat v}} 
d'_m d'_{m\xor v} \\
&=&\frac{2}{\alpha}\sum_{m\in\Gamma_{\hat v}}
(\alpha d_{m\xor v}')(d_m')\\
&\le&\alpha\sum_{m\in\Gamma_{\hat v}}
d_{m\xor v}'^2
+\frac{1}{\alpha}\sum_{m\in\Gamma_{\hat v}}d_m'^2\\
&\le&\alpha\sum_{\stackrel{m\in S^c_{\mathbf s},n\in S_{\mathbf s}}
{|m\xor n|\ge\frac{\hat v}{2}}}
 d_{m\xor n\xor v}^2+\frac{1}{\alpha}
\sum_{\stackrel{m\in S^c_{\mathbf s},n\in S_{\mathbf s}}
{|m\xor n|\ge\frac{\hat v}{2}}}
d_{m\xor n }^2 \\
&=&\alpha\sum_{|l|\ge\hat \frac{v}{2}} d_{l\xor v}^2+\frac{1}{\alpha}
\sum_{|l|\ge\frac{\hat v}{2}}d_l^2
\end{eqnarray*}
Due to the fact that the $d_l^2$ form a probability distribution,
any sum of them is less than or equal to unity.
\begin{equation}
SD_v\le\alpha+\frac{1}{\alpha}\sum_{|l|\ge\frac{\hat v}{2}}d_l^2
\end{equation}
\mbox{ } \hfill {\bf QED}

Note that the 
number of parity check strings $r$ doesn't appear in the
final expression, and this might seem surprising.
However, it does appear there implicitly, since increasing $r$ by one
increases the number of parity check strings from $2^r - 1$ to 
$2^{r+1} - 1$, hence potentially decreases $\hat{v}$.

\section{Security of the Entire Key}\label{app:m-bits}

We give a proof that bitwise security implies security of the entire string.
This is first shown classically, and then making use of
Shannon Distinguishability, the same bound holds for quantum bits.

\subsection{Classical Information Theory}
\begin{lemm}
For independent random variables ${\cal A}_i$, $i \in (1,2,\ldots ,m)$
and random variable ${\cal E}$\\ 
$I({\cal A}_i;{\cal E}|{\cal A}_1,{\cal A}_2,\ldots ,{\cal A}_{i-1})
\le 
I({\cal A}_i;{\cal E}|{\cal A}_1,{\cal A}_2,\ldots ,{\cal A}_{i-1},{\cal A}_{i+1},\ldots,{\cal A}_{m})$
\end{lemm}
{\bf Proof:}
%
%
%
First we define a few sets: ${\cal A}_{<i}\equiv 
\{{\cal A}_1,{\cal A}_2,\ldots ,{\cal A}_{i-1}\}, \
{\cal A}_{>i}\equiv 
\{{\cal A}_{i+1},{\cal A}_{i+2},\ldots ,{\cal A}_{m}\},$ and
${\cal A}_{\ne i}\equiv 
\{{\cal A}_{1},{\cal A}_{2},\ldots ,{\cal A}_{i-1},{\cal A}_{i+1},\ldots ,
{\cal A}_{m}\}$.  Of course, ${\cal A}_{\ne i}={\cal A}_{< i}\cup 
{\cal A}_{>i}$.  In this notation the lemma says:
$
I({\cal A}_i;{\cal E}|{\cal A}_{\ne i})-
I({\cal A}_i;{\cal E}|{\cal A}_{< i})\ge 0
$

\begin{eqnarray*}
I({\cal A}_i;{\cal E}|{\cal A}_{\ne i})-
I({\cal A}_i;{\cal E}|{\cal A}_{< i})&=&
H({\cal A}_i|{\cal A}_{\ne i})-
H({\cal A}_i|{\cal E},{\cal A}_{\ne i})
-H({\cal A}_i|{\cal A}_{< i})+
H({\cal A}_i|{\cal E},{\cal A}_{< i}) \\
&=&(H({\cal A}_i|{\cal E},{\cal A}_{< i})
-H({\cal A}_i|{\cal E},{\cal A}_{\ne i}))
-(H({\cal A}_i|{\cal A}_{< i})
-H({\cal A}_i|{\cal A}_{\ne i})) \\
&=&(H({\cal A}_i|{\cal E},{\cal A}_{< i})
-H({\cal A}_i|{\cal E},{\cal A}_{< i},{\cal A}_{> i})) \\
&\ \ \ \ \ \ \ \ -&(H({\cal A}_i|{\cal A}_{< i})
-H({\cal A}_i|{\cal A}_{< i},{\cal A}_{> i})) \\
&=&I({\cal A}_i;{\cal A}_{> i}|{\cal E},{\cal A}_{< i})
-I({\cal A}_i;{\cal A}_{> i}|{\cal A}_{< i})
\end{eqnarray*}

Due to the independence of ${\cal A}_i$, 
$I({\cal A}_i;{\cal A}_{> i}|{\cal A}_{< i})=0$.  Since any information
is non-negative,  
$I({\cal A}_i;{\cal A}_{> i}|{\cal E},{\cal A}_{< i})\ge 0$.  Hence
$
I({\cal A}_i;{\cal A}_{> i}|{\cal E},{\cal A}_{< i})-
I({\cal A}_i;{\cal A}_{> i}|{\cal A}_{< i})\ge 0
$
\hfill {\bf QED}
\begin{thm}
For independent random variables ${\cal A}_i$, $i \in (1,2,\ldots ,m)$
and random variable ${\cal E}$\\
$I({\cal A}_1,{\cal A}_2,\ldots ,{\cal A}_m;{\cal E})\le m\ max_i(
I({\cal A}_i;{\cal E}|{\cal A}_1,{\cal A}_2,\ldots ,
{\cal A}_{i-1},{\cal A}_{i+1},\ldots,{\cal A}_{m}))$
\end{thm}
{\bf Proof:}
Here we simply apply the chain rule for mutual information\cite{CoverThomas}
and we then apply the above lemma.  We will use the same notions
introduced in the previous proof.

\begin{eqnarray*}
I({\cal A}_1,{\cal A}_2,\ldots ,{\cal A}_m;{\cal E})&=&
\sum_i I({\cal A}_i;{\cal E}|{\cal A}_{<i}) \\
&\le& \sum_k I({\cal A}_k;{\cal E}|{\cal A}_{\ne k}) \\
&\le& \sum_k max_i(I({\cal A}_i;{\cal E}|{\cal A}_{\ne i}))\\
&=&m\ max_i(I({\cal A}_i;{\cal E}|{\cal A}_{\ne i}))\\
\end{eqnarray*}

\hfill {\bf QED}

\begin{lemm}
For independent random variables ${\cal A}_i$, $i \in (1,2,\ldots ,m)$
and random variable ${\cal E}$\\
$I({\cal A}_1,{\cal A}_2,\ldots ,{\cal A}_m;{\cal E})\le m\ max_{i,a_{\ne i}}
(I({\cal A}_i;{\cal E}|{\cal A}_{\ne i}=a_{\ne i}))$. 
Where $a_{\ne i}$ is a set of outcomes for all ${\cal A}$ except $i$.
\end{lemm}
{\bf Proof:}
We must simply prove $I({\cal A}_i;{\cal E}|{\cal A}_{\ne i})\le
max_{a_{\ne i}}I({\cal A}_i;{\cal E}|{\cal A}_{\ne i}=a_{\ne i})$ and
then apply the previous theorem.
\begin{eqnarray*}
I({\cal A}_i;{\cal E}|{\cal A}_{\ne i})&=&
\sum_{a_{\ne i}}P({\cal A}_{\ne i}=a_{\ne i})
I({\cal A}_i;{\cal E}|{\cal A}_{\ne i}=a_{\ne i}) \\
&\le&\sum_{a_{\ne i}}P({\cal A}_{\ne i}=a_{\ne i}) max_{a'_{\ne i}}
I({\cal A}_i;{\cal E}|{\cal A}_{\ne i}=a'_{\ne i}) \\
&=&max_{a_{\ne i}}I({\cal A}_i;{\cal E}|{\cal A}_{\ne i}=a_{\ne i})
\end{eqnarray*}
\hfill {\bf QED}

\subsection{Quantum Connection}
We have used classical information theory to prove the above identities.
In the quantum setting, Eve has a quantum system that may depend
on Alice's bits, ${\cal A}_i$.
The classical formulas are all valid once a particular measurement 
on the system (POVM)
is fixed by Eve, so that:
\begin{equation}
I({\cal A}_1,{\cal A}_2,\ldots ,{\cal A}_m;{\cal E}^M)\le m\ 
max_{i,a_{\ne i}}
I({\cal A}_i;{\cal E}^M|{\cal A}_{\ne i}=a_{\ne i})
\end{equation}
where ${\cal E}^M$ is the random variable
obtained by Eve's output from her measurement $M$.
In particular the above is true for any
measurement, $\tilde{M}$, that Eve may consider optimal
to learn
the bits of Alice's key, ${\cal A}_i$, all at once.
 
Now we need the definition of Shannon Distinguishability:
\begin{equation}
SD^{i,a_{\ne i}}
\equiv sup_M I({\cal A}_i;{\cal E}^M|{\cal A}_{\ne i}=a_{\ne i})
\end{equation}
Note, a measurement that achieves (or nearly achieves) this upper bound
may not be optimal for eavesdropping on the entire key, but that is of no
consequence to the proof.
Therefore,
$I({\cal A}_i;{\cal E}^M|{\cal A}_{\ne i}=a_{\ne i})\le SD^{i,a_{\ne i}}$ 
for all $M$ and in
particular
\begin{equation}
I({\cal A}_i;{\cal E}^{\tilde{M}}|{\cal A}_{\ne i}=a_{\ne i})\le SD^{i,a_{\ne i}}
\end{equation} 
Hence we have a bound for total mutual information for
any measurement Eve might consider optimal:
\begin{equation}
I({\cal A}_1,{\cal A}_2,\ldots ,{\cal A}_m;{\cal E}^{\tilde{M}})
\le m\ max_{i,a_{\ne i}} SD^{i,a_{\ne i}}
\end{equation}

\section{A Proof of Lemma \em \ref{sec1-lemm}}
\label{exp-bound1}

To prove this lemma we first note that
$P(c_T,i_T,{\cal T}={\rm pass}|b,s ) = P(c_T,i_T|b,s) $ when
we sum only over such terms $c_T$
that pass the test.  Then we apply equation \ref{eq:inf}: 
\begin{eqnarray*}
I({\cal A};{\cal E}|i_T,j_T,b,s) \le
m \left[
\alpha+\frac{1}{\alpha}\sum_{|c_I|\ge\frac{\hat v}{2}}
P(c_{I}^o|i_T,j_T,b,s)\ \right]
\end{eqnarray*}

Thus:
\begin{eqnarray*}
\sum_{i_T,c_T}
P({\cal T}=pass,i_T,c_T|b,s)I({\cal A};{\cal E}|i_T, c_T, b,s)
&=&
	\sum_{\frac{|c_T|}{n}\le p_{allowed}}\sum_{i_T}
P(i_T,c_T|b,s)I({\cal A};{\cal E}|i_T, c_T, b,s) \\
&\hspace*{-6in}\le&\hspace*{-3in}
m\left(\alpha\sum_{\frac{|c_T|}{n}\le p_{allowed}}
\sum_{i_T}P(i_T,c_T|b,s)+\frac{1}{\alpha}
\sum_{\frac{|c_T|}{n}\le p_{allowed}}\sum_{i_T}P(i_T,c_T|b,s)
 \sum_{|c_I| >\frac{\hat{v}}{2}} P(c_I^o|c_T,i_T,b,s)   \right)
\\
&\hspace*{-6in}=&\hspace*{-3in}
m\left(\alpha P({\cal T}=pass \ |\ b,s)+\frac{1}{\alpha}
\sum_{\frac{|c_T|}{n}\le p_{allowed}}\sum_{|c_I| >\frac{\hat{v}}{2}}
\sum_{i_T}
P(c_I^o,c_T,i_T|b,s)\right) \\
&\hspace*{-6in}=&\hspace*{-3in}
m\left(\alpha P({\cal T}=pass \ |\ b,s)+\frac{1}{\alpha}
\sum_{\frac{|c_T|}{n}\le p_{allowed}}\sum_{|c_I| >\frac{\hat{v}}{2}}
P(c_I^o,c_T|b,s)\right)
\end{eqnarray*}
The dependence on the $i_T$ has been 
removed by averaging over all values.
Since, the probability of
passing the test, $P({\cal T}=pass |\ b,s)$ is less than unity we replace
it by $1$.
Also, \\
$\sum_{\frac{|c_T|}{n}\le p_{allowed}}\sum_{|c_I| >\frac{\hat{v}}{2}}
P(c_I^o,c_T|b,s) \equiv 
P[(\frac{|C_I^o|}{n}>\frac{\hat{v}}{2})    \cap
(\frac{|C_T|}{n}\le p_{allowed})|b,s]$
Thus,
\begin{eqnarray*}
\sum_{i_T,c_T}
P({\cal T}=pass,i_T,c_T|b,s)I({\cal A};{\cal E}|i_T, c_T, b,s)
&\le&m\left(
\alpha+\frac{1}{\alpha}
P[(\frac{|C_I^o|}{n}>\frac{\hat{v}}{2}) \cap
(\frac{|C_T|}{n}\le p_{allowed})|b,s]\right)
\end{eqnarray*}


\section{A Proof of Lemma \em \ref{sec1-lemm2}}
\label{exp-bound2}

Here we show that:
\begin{equation}
\label{eq:basis_ave}
\sum_b P\left((\frac{|C_I^o|}{n}>p_{a}+\epsilon) \cap
(\frac{|C_T|}{n}\le p_{a})|b,s\right)=
\sum_b
P\left((\frac{|C_I|}{n}>p_{a}+\epsilon) \cap
(\frac{|C_T|}{n}\le p_{a})|b,s\right)
\end{equation}
There are $2^{2n}$ possible basis strings that Alice is equally likely to
choose.  Therefore $P(b)=\frac{1}{2^{2n}}$.
Combining the above with the average over $b$ of lemma \ref{sec1-lemm}
we get the desired result:
\begin{equation}
\sum_{i_T,c_T,b}
P({\cal T}=pass,i_T,c_T,b|s)I({\cal A};{\cal E}|i_T, c_T, b,s)
\le \alpha + \frac{1}{\alpha}
	  \frac{1}{2^{2n}}\sum_b P[(\frac{|C_I|}{n}>p_{a}+\epsilon) \cap
	(\frac{|C_T|}{n}\le p_{a})|b,s]
\end{equation}

We have assumed a particular basis string $b$ was announced, and
that a bit-selection string $s$ was announced.
For a fixed bit-selection $s$ and bases $b$ we define the unique
strings $b_I$ and $b_T$ which are two substrings 
of $b$ where one contains
only the bases of the information bits, and the other only the bases of the
test bits, while the internal order (inside $b_I$ and $b_T$ is as the
internal order in $b$.
For instance, if $s = 111100001100$, and $b = 010100110101$ then
$b_I = 001101$ and $b_T = 010101$.

Recall that $C_I^o$ is the error
string had the \emph{opposite} basis been used for the information bits.
Thus we can equivalently write:
\begin{eqnarray*} 
P\left((\frac{|C_I^o|}{n}>p_{a}+\epsilon) \cap
(\frac{|C_T|}{n}\le p_{a})|b_I,b_T,s\right)
&\equiv&P\left((\frac{|C_I|}{n}>p_{a}+\epsilon) \cap
(\frac{|C_T|}{n}\le p_{a})|\bar{b_I},b_T,s\right)
\end{eqnarray*} 

Averaging over all bases of the information bits $b_I$ we get
\begin{eqnarray*}
\sum_{b_I}P\left((\frac{|C_I|}{n}>p_{a}+\epsilon) \cap
(\frac{|C_T|}{n}\le p_{a})|\bar{b_I},b_T,s\right)
&=&\sum_{b_I}P\left((\frac{|C_I|}{n}>p_{a}+\epsilon) \cap
(\frac{|C_T|}{n}\le p_{a})|b_I,b_T,s\right) \ ,
\end{eqnarray*}

Now we can also sum over the bases of the test bits to get
\begin{eqnarray*}
\sum_{b_I,b_T}P\left((\frac{|C_I|}{n}>p_{a}+\epsilon) \cap
(\frac{|C_T|}{n}\le p_{a})|\bar{b_I},b_T,s\right)
&=&\sum_{b_I,b_T}P\left((\frac{|C_I|}{n}>p_{a}+\epsilon) \cap
(\frac{|C_T|}{n}\le p_{a})|b_I,b_T,s\right) \ ,
\end{eqnarray*}
so that
\begin{equation}
\sum_{i_T,c_T,b}
P({\cal T}=pass,i_T,c_T,b|s)I({\cal A};{\cal E}|i_T, c_T, b,s)
\le \alpha + \frac{1}{\alpha}
	  \frac{1}{2^{2n}}\sum_b P[(\frac{|C_I|}{n}>p_{a}+\epsilon) \cap
	(\frac{|C_T|}{n}\le p_{a})|b,s]
\end{equation}

Note that in the above equation 
the dependence on $b$ in the right hand side
has been removed by averaging.  This is just the average 
over all choices of
basis.

\section{A Proof of Lemma \em \ref{sec1-lemm3}}
\label{exp-bound3}

To prove that:
\begin{equation}
\sum_{i_T,c_T,b,s}
P({\cal T}=pass,i_T,c_T,b,s)I({\cal A};{\cal E}|i_T, c_T, b,s)
\le 2m\sqrt{\frac{1}{2^{2n}}\sum_b 
P[(\frac{|C_I|}{n}>p_a+\epsilon) \cap
(\frac{|C_T|}{n}\le p_a)|b]}
\end{equation}
we start with:

\begin{eqnarray*}
\sum_{i_T,c_T,b}P({\cal T}={\rm pass},i_T,c_T,b\ |\ s)
I({\cal A};{\cal E}\ |\ i_T,c_T,b,s)
\le m\left(\alpha + \frac{1}{\alpha}
	 \frac{1}{2^{2n}}\sum_b P[(\frac{|C_I|}{n}>p_{a}+\epsilon) \cap
	(\frac{|C_T|}{n}\le p_{a})|b,s]\right) \ .
\end{eqnarray*}

Each order string $s$ must have an equal number of zeros and ones.  Aside
from that, each are equally likely.  So Alice chooses them with 
$P(s)=\frac{1}{{2n \choose n}}$.  The right
hand side of the average over $s$ of the above equation is:

\begin{eqnarray*}
m\left(\alpha + \frac{1}{\alpha}
	 \frac{1}{2^{2n}}\sum_b\left[\frac{1}{{2n \choose n}}\sum_s 
P[(\frac{|C_I|}{n}>p_{a}+\epsilon) \cap
	(\frac{|C_T|}{n}\le p_{a})|b,s]\right]\right)
\end{eqnarray*}

We are now, for each fixed bases $b$, 
able to bound each term in the $[]$ parenthesis 
by a law of large numbers.
The probability that a test string will be chosen such that it passes
the test, but the remaining information string would not pass the test,
given basis $b$, is written as:
\begin{eqnarray*}
P\left((\frac{|C_I|}{n}>p_{a}+\epsilon) \cap
(\frac{|C_T|}{n}\le p_{a})|b\right)
&\equiv&\frac{1}{{{2n}\choose {n}}}\sum_s
P\left((\frac{|C_I|}{n}>p_{a}+\epsilon) \cap
(\frac{|C_T|}{n}\le p_{a})|b,s\right)
\end{eqnarray*}
Making use of the above and choosing 
$\alpha=\frac{1}{2^{2n}}\sum_b P[(\frac{|C_I|}{n}>p_a+\epsilon) \cap
(\frac{|C_T|}{n}\le p_a)|b]$ to minimize the
bound we have the result.
Hence,
\begin{equation}
\sum_{i_T,c_T,b,s}
P({\cal T}=pass,i_T,c_T,b,s)I({\cal A};{\cal E}|i_T, c_T, b,s)
\le
	 2m\sqrt{\frac{1}{2^{2n}}\sum_b 
P[(\frac{|C_I|}{n}>p_a+\epsilon) \cap
	(\frac{|C_T|}{n}\le p_a)|b]}
\end{equation}
\mbox{ } \hfill {\bf QED}

\section{A Proof of Lemma \em \ref{sec2-lemm}}
\label{law-large-num}

In this Appendix let $c$ denote the combination of $c_T$ and $c_I$,
instead of the combination of $c_T$, $c_U$ and $c_I$, and let $C$ be
the random variable corresponding to $c$.
Let
\begin{eqnarray*}
 h_b &\equiv& P[(\frac{|C_I|}{n}>p_{allowed}+\epsilon)\cap
 (\frac{|C_T|}{n}<p_{allowed} )|b]\\
\end{eqnarray*}
This $h_b$ is the probability that the information has $\epsilon$ more
than the allowed error rate, when the test has less than the allowed error
rate, averaged over all choices of test and information, for a particular
basis b.

\begin{eqnarray*}
h_b&=&
\sum_{|c|}P[(\frac{|C_I|}{n}>p_{allowed}+
\epsilon) \cap (\frac{|C_T|}{n} \le p_{allowed}) \cap (|C|=|c|)|Basis=b] \\
&=&\sum_{|c|}P[ \{ (\frac{|C_I|}{n}>p_{allowed}+\epsilon) \cap
(\frac{|C_T|}{n} \le p_{allowed}) \}\ {\rm given} \  |c|,b] 
P(|c| \ {\rm given} \ b)
\end{eqnarray*}
Note that in principle, $P(c|b)$, and hence also 
$P(|c| \ {\rm given} \ b)=P(|C|=|c| \ {\rm given}\ b)$ can
be calculated,
but we shall soon see that there is no need to
calculate $P(|c|\ {\rm given} \ b)$.

Now we must note that
$P \left[ \{ (\frac{|C_I|}{n}>p+\epsilon) \cap
(\frac{|C_T|}{n} \le p) \} \ {\rm given} \ |c|,b\right]$,
does not
depend on the attack.  And in fact, in the aforementioned
equation, the basis $b$ is superfluous.  Once a basis is fixed,
and the numbers of errors given ($|c|$ and $b$), then we are
safely in the hands of random samplings.  Any sample choice,
of bits for use as information bits will not change $|c|$ or
$b$.
Of course $c_I$, $c_T$ and $c$ are not independent.
By definition, $|c|=|c_I|+|c_T|$.
If $|c_I| > n(p_{allowed}+\epsilon)$ and $|c_T| \le n p_{allowed}$,
then $|c_I|-|c_T| > n\epsilon$.  So:
\begin{eqnarray*}
 P[ \{ (|c_I|>n(p_{allowed}+\epsilon) ) \cap (|c_T|<n p_{allowed}) \}
\  {\rm given}\  |c|,b]
 &\le&P[(|c_I|> \frac{|c|}{2}+\frac{n\epsilon}{2})\  {\rm given}\  |c|,b]
\end{eqnarray*}
It is the probability that a sampled subset has a
weight which is $n\epsilon/2$
more than the average.
Intuitively, it may be obvious that the weight of the test
string should be equal to half the weight of the full string.
By Hoeffding's bound\footnote{BTW -
A factor of 2 can be improved.}
we make this rigorous (see Appendix \ref{app:Hoeff})
\begin{equation}
P[(|c_I| > \frac{|c|}{2}+\frac{n\epsilon}{2}) \ {\rm given} \ |c|,b]
\le 2 e^{-\frac{1}{2}n\epsilon^2}
\ .
\end{equation}

Hence
\begin{eqnarray*}
h_b &=& 
 P(\frac{|C_I|}{n}>p_{allowed}+\epsilon,\frac{|C_T|}{n}<p_{allowed}|Basis=b )\\
&=&\sum_{|c|}P(\frac{|C_I|}{n}>p_{allowed}+\epsilon,
\frac{|C_T|}{n} \le p_{allowed},\ {\rm given} \ |c|,b)P(|c| \ {\rm given} \ b) \\
&\le&
 2 e^{-\frac{1}{2} n \epsilon^2} 
\sum_{|c|}P(|c| \ {\rm given} \ b) \\
&=&
2 e^{-\frac{1}{2} n \epsilon^2}
\end{eqnarray*}

So finally we summarize the result to be
\begin{equation} \label{Hoefding2}
h_b =
  P(\frac{|C_I|}{n}>p_{allowed}+\epsilon,\frac{|C_T|}{n}<p_{allowed}|Basis=b)
\le 2 e^{-\frac{1}{2}n\epsilon^2}
\ .
\end{equation}

This result for $h$ is useful both for the reliability proof and the
security proof. \hfill {\bf QED}

\section{Hoeffding}
\label{app:Hoeff}

We need to bound the probability
$P(|c_I|>\frac{|c|}{2}+\frac{n\epsilon}{2}, \ {\rm given} \  |c|)$.
Recall that half of the bits are randomly selected to be test bits.
This is
random sampling without replacement.
For the above probability we are given
an error string $c$.
Each bit in the error string is either zero or one depending
on whether there is or is not an error respectively.

Therefore,
$P(|c_I|>\frac{|c|}{2}+\frac{n\epsilon}{2},\ {\rm given}\  |c|)$
is the probability that
a sampled average is greater than the entire
sample average by more than $\epsilon/2$.  This
case was studied by Hoeffding~\cite{Hoeffding}.
The following bound is given in \cite{DGL}
\begin{equation}
P(|\frac{|c_I|}{n}-\frac{|c|}{2n}|\ge\epsilon/2)\le
2e^{-\frac{n\epsilon^2}{2}}
\end{equation}

Note that $A$ in \cite{DGL} represents $c$ in our notations,
$m$ is the sample average $|c|/2n$,
and the sum of $Z_i$ is $|c_I|$.

Of course this bound is more restrictive than we need (due to the use of
the absolute value).  We only need
$P(\frac{|c_I|}{n}-\frac{|c|}{2n}\ge\epsilon/2)$,
which is smaller than the above bound
and therefore the above bound suffices.
In fact, from Hoeffding's original
paper~\cite{Hoeffding} we can get the bound:
\begin{equation}
P(\frac{|c_I|}{n}\ge\frac{|c|}{2n}+\epsilon/2)\le
e^{-\frac{n\epsilon^2}{2}}
\end{equation}

\section{Satisfying the Security Criterion}
\label{app:sec_crit}

So far we have not shown that the security criterion is satisfied
by bounding the following:

\begin{equation}
\sum_{i_T,c_T,b,s}
P({\cal T}=pass,i_T,c_T,b,s)I({\cal A};{\cal E}|i_T, c_T, b,s)
\leq e^{2(\alpha-\beta n)}
\end{equation} 
We now show
that when the above bound is satisfied, as shown in the paper, then
the security criterion is satisfied:
\begin{equation}
Prob({\rm Test\ Passes\ and\ } I_{Eve} \ge
e^{\alpha-\beta n})\leq e^{\alpha-\beta n}
\end{equation}
Where $I_{Eve}\equiv I({\cal A};{\cal E}|i_T, c_T, b,s)$.

To show the above break the sum into the parts
where Eve has large information and the part where she has small.  Then
standard bounding techniques are used:
\begin{eqnarray*}
\sum_{i_T,c_T,b,s}P({\cal T}=pass,i_T,c_T,b,s)
I({\cal A};{\cal E}|i_T, c_T, b,s)&=&
\sum_{\stackrel{i_T,c_T,b,s}{s.t. \ I_{Eve}<I'}}
P({\cal T}=pass,i_T,c_T,b,s)
I({\cal A};{\cal E}|i_T, c_T, b,s) \\
&+&\sum_{\stackrel{i_T,c_T,b,s}{s.t. \ I_{Eve}\ge I'}}
P({\cal T}=pass,i_T,c_T,b,s)
I({\cal A};{\cal E}|i_T, c_T, b,s) \\
&\ge&\sum_{\stackrel{i_T,c_T,b,s}{s.t. \ I_{Eve}\ge I'}}
P({\cal T}=pass,i_T,c_T,b,s)
I({\cal A};{\cal E}|i_T, c_T, b,s) \\
&\ge&\left(\sum_{\stackrel{i_T,c_T,b,s}{s.t. \ I_{Eve}\ge I'}}
P({\cal T}=pass,i_T,c_T,b,s)\right)I' \\
\end{eqnarray*}
The above steps follow from non-negativity of probability and mutual 
information.  We are really already done:
\begin{equation}
\left(\sum_{\stackrel{i_T,c_T,b,s}{s.t. \ I_{Eve}\ge I'}}
P({\cal T}=pass,i_T,c_T,b,s)\right)I' \leq 
\sum_{i_T,c_T,b,s}P({\cal T}=pass,i_T,c_T,b,s)
I({\cal A};{\cal E}|i_T, c_T, b,s)
\end{equation}
So far $I'$ is a free parameter.  We can set it to any value we like, namely \\
$I'=\sqrt{ \sum_{i_T,c_T,b,s}P({\cal T}=pass,i_T,c_T,b,s)
I({\cal A};{\cal E}|i_T, c_T, b,s)  }$:
\begin{eqnarray*}
Prob({\rm Test\ Passes\ and\ } I_{Eve} \ge I')&=&
\sum_{\stackrel{i_T,c_T,b,s}{s.t. \ I_{Eve}\ge I'}} 
P({\cal T}=pass,i_T,c_T,b,s)  \\
&\leq&\sqrt{ \sum_{i_T,c_T,b,s}P({\cal T}=pass,i_T,c_T,b,s)
I({\cal A};{\cal E}|i_T, c_T, b,s)  } \\
\end{eqnarray*}
If we assume that $\sum_{i_T,c_T,b,s}P({\cal T}=pass,i_T,c_T,b,s)
I({\cal A};{\cal E}|i_T, c_T, b,s)
\leq e^{2(\alpha-\beta n)}$ then we have:
\begin{equation}
Prob({\rm Test\ Passes\ and\ } I_{Eve} \ge e^{\alpha-\beta n})
\leq e^{\alpha-\beta n}
\end{equation}
Thus, the bounds that we have shown satisfy the security criterion.

\section{Existence of Codes for Both Reliability and Security}
\label{app:codes-exist}

Choosing a code which is good
when $n$ is large (for constant error rate) is
not a trivial problem in ECC. A Random Linear Code (RLC)
is one such code, however, it does not
promise us that the distances are as required,
but only gives the desired
distances with probability as close to one as we want.
With RLC, we find that the threshold below which a secure key
can be obtained is $p_{allowed} \le 7.56 \%$.

In order to correct $t$ errors with certainty,
a code must have a minimal
Hamming distance between the code words $d \ge 2t +1$ so that all
original code
words, even when distorted by $t$ errors, can still be identified
correctly.
For any $c_T$ which passes the test,
we are promised (due to Lemma~\ref{sec2-lemm})
that the probability of having
$t = |c_I| > n(p_{allowed} + \epsilon_{\rm rel}) $
errors is smaller than
$ h = 2 e^{-(1/2) n \epsilon_{\rm rel}^2}$.

Thus, we need to choose a RLC that promises a Hamming distance at
least $d$ such that
$p_{allowed} + \epsilon_{\rm rel} <   t/n = \frac{d-1}{2n}$,
and then the
$t$ errors are corrected except for a probability
smaller than $ h_1 =  2 e^{-(1/2) n \epsilon_{\rm rel}^2}$.

For any $n,r=n-k$, and for $\delta$ such that $H_2(\delta)< r/n$,
an arbitrary {\it random linear code} $(n,k,d)$
satisfies $d/n\ge \delta$,
except for a probability (see~\cite{Gallager},
Theorem 2.2)
\< \hbox{Prob}(d/n<\delta) \le
	\frac{c(\delta)}{\sqrt{n}} 2^{n(H_2(\delta)-r/n)} = g_1 \>
where $c(\delta) = \frac{1}{1-2\delta}
\sqrt{\frac{1-\delta}{2\pi\delta}}$.

If we choose $\delta = 2(p_{allowed} + \epsilon_{\rm rel}) + 1/n$
then we are promised that the errors are corrected,
except for probability
that the error rate is larger than expected or a bad code was chosen.

Using such a code,
$\epsilon_{\rm rel}$ is now a function of $\delta$ so that
$ \epsilon_{\rm rel} = \delta/2 - 1/(2n) - p_{allowed}$ and therefore,
\begin{eqnarray}
h_1 =  2 e^{-(n/4)(\delta-\frac{1}{n}-2p_{allowed})^2}
\end{eqnarray}
and almost all such codes correct all the errors.

Therefore, the code is reliable except for a probability
$g_1 + h_1$.

The above result can be improved~\cite{Mayers98}
by taking RLC with distance
$d-1 \ge n(p_{allowed} +
\epsilon_{\rm rel})$ (without the factor of 2),
since such a code can also correct
$ t = n(p_{allowed} + \epsilon_{\rm rel})$ errors except for an
exponentially small fraction $f_1$ of the possible errors.
We get 
\begin{eqnarray}
f_1 =  2 e^{-(n/4)(\delta-\frac{1}{n}-p_{allowed})^2}
\end{eqnarray}
and it is exponentially small (in the limit of large $n$) for any
$\delta > p_{allowed}$.

Recall that we choose $\epsilon_{\rm sec}$ such that
$|v|\ge 2n(p_{allowed}+\epsilon_{\rm sec})$.
Let $|v|$ be the minimal distance between one PA string and any other
parity check string (or linear combination) taken from ECC and PA.
Clearly, the Hamming weight of the dual code of the ECC, once the PA is
also added, provides a lower bound on $|v|$.
Thus, it is sufficient to demand
$d^\perp \ge 2n(p_{allowed}+\epsilon_{\rm sec})$ in order to prove
security.
Choosing a RLC for the ECC and PA,
one cannot be completely sure that the distance
indeed satisfies the constraint, but this shall be true with probability
exponentially close to one.
We use the dual code $(n,r^\perp,d^\perp)$, where $r^\perp=n-r-m$.
Such codes satisfy $d^\perp/n\ge \delta^\perp$, except for a fraction of
\< \hbox{Prob}(d^\perp/n<\delta^\perp) \le
\frac{c(\delta^\perp)}{\sqrt{n}} 2^{n(H_2(\delta^\perp)-(n-r-m)/n)}
= g_2 \>

With $\delta^\perp = 2(p_{allowed}+\epsilon_{\rm sec})$.

Assuming that Eve gets full information when the code
fails we get:
\begin{equation}
\sum_{i_T,c_T,b,s} P({\cal T}=pass,i_T,c_T,b,s)
I({\cal A};{\cal E}|i_T,c_T,b,s)
\le m\left(2\sqrt{2e^{-\frac{1}{2}n\epsilon_{\rm sec}^2}}+g_2\right)
\end{equation}
Since the first term is exponentially small we only need
look at $g_2$.   We also need to worry about the reliability so we
need $g_1 $ and $f_1$ to be exponentially small as well.
All of them are exponentially small if the following
conditions are met:
\begin{eqnarray*}
H_2(\delta)-r/n&<&0 \\
H_2(\delta^{\perp})+r/n+m/n-1&<&0 \\
\end{eqnarray*}
Or written another way:
\begin{eqnarray*}
H_2(p_{allowed}+\epsilon_{\rm rel}+1/n) &<&r/n \\
H_2(2p_{allowed}+2\epsilon_{\rm sec})+
H_2(p_{allowed}+\epsilon_{\rm rel}+1/n)
&<&1-R_{secret}
\end{eqnarray*}

Where $R_{secret}\equiv m/n$.
In the limit of large $n$ and $\epsilon$'s close to zero,
$p_{allowed}<7.56\%$
satisfies the bound and hence this is our threshold.

Asymptotically, any $R_{secret}<1-H_2(2p_a)-H_2(p_a)$ is secure
and reliable for the given ECC+PA.  Note, as $p_a$ goes to zero,
$R_{secret}$ goes to $1$, which means all the information bits are
secret.

This threshold is based on the property of the code, and other codes
might give worse thresholds. It is possible to replace the RLC by a
code that can be decoded and encoded efficiently (e.g., Reed-Solomon
concatenated code), and add random PA strings. The Hamming distance
between the PA check-strings and the ECC check-strings is still
bounded below in the same way as for the RLC (see~\cite{Mayers98}).

A better threshold can be obtained by using privacy-distillation instead
of the standard ECC+PA approach.

Note that any probability of failure in the classical transmission can
be added in the same way that $g_2$ is added. This is important to
prove security in the case where a fault-tolerant
classical transmission is not 100\%
reliable. It shows an important advantage over the
proof of~\cite{LC98} which is based on fault tolerant quantum ECC.

\end{document}